\documentclass[12pt,a4paper]{article}
\usepackage[utf8]{inputenc}
\usepackage{bm}
\usepackage{color}
\usepackage{amsmath}
\usepackage{amsthm}
\usepackage{amssymb}
\usepackage{graphicx}
\usepackage{verbatim}
\usepackage[breaklinks]{hyperref}

\newtheorem{lemma}{Lemma}
\newtheorem{theorem}{Theorem}
\newtheorem{corollary}{Corollary}

\theoremstyle{definition}
\newtheorem{definition}{Definition}

\newtheorem{oldtheorem}{Theorem}

\newcommand{\set}[2]{\ensuremath{ \{ \, #1 \mid #2 \, \} }}
\newcommand{\setbig}[2]{\big\{ \: #1 \;\big|\; #2 \: \big\}}

\renewcommand{\epsilon}{\varepsilon}

\begin{document}
\sloppy

\title{Complexity of the emptiness problem for graph-walking automata and for tilings with star subgraphs\thanks{%
		This work was supported by the Russian Science Foundation, project 18-11-00100.}}
\author{Olga Martynova\thanks{
	Department of Mathematics and Computer Science,
	St.~Petersburg State University,
	7/9 Universitetskaya nab., Saint Petersburg 199034, Russia,
	\texttt{olga22mart@gmail.com}.
}}

\maketitle

\begin{abstract}
This paper proves the decidability of the emptiness problem for two models which recognize graphs:
graph-walking automata,
and tilings of graphs by star subgraphs (star automata).
Furthermore, it is proved that the non-emptiness problem for graph-walking automata
(that is, whether a given automaton accepts at least one graph)
is NEXP-complete.
For star automata, which generalize nondeterministic tree automata to the case of graphs,
it is proved that their non-emptiness problem is NP-complete.
\end{abstract}

\section{Introduction}

The main result of this paper
is the decidability of the emptiness problem for graph-walking automata
and its computational complexity.

A graph-walking automaton is a model of a robot in a maze.
It has finitely many states, and it deterministically walks on graphs
with labelled nodes and labelled edge end-points.
The automaton decides by which edge to move
depending on the label of the current node and on its current state.
The automaton can also decide to accept or to reject,
and so it defines a graph language: the set of graphs it accepts.

Graph-walking automata were first introduced by Michael Rabin,
who stated the conjecture
that for each graph-walking automaton,
even if it is additionally allowed to use finitely many pebbles,
there is a graph that it cannot fully explore.
Budach~\cite{Budach} proved this conjecture for graph-walking automata without pebbles.
Later Fraigniaud et al.~\cite{FraigniaudIlcinkasPeerPelcPeleg}
gave an easier proof of this fact.
Rollik~\cite{Rollik} proved that not only pebbles,
but even co-operation of several interacting automata
would not help to traverse every graph,
thus proving Rabin's conjecture.
Kunc and Okhotin~\cite{KuncOkhotin_reversible} showed
that every graph-walking automaton can be transformed
to an automaton which halts on every input,
to an automaton which accepts only at the initial node,
and to a reversible automaton,
which all accept the same set of graphs.
Later Martynova and Okhotin~\cite{MartynovaOkhotin_lb}
reduced the number of states
needed for these transformations,
and obtained asymptotically tight lower bounds.

Overall, graph-walking automata have been studied for a long time,
and it is natural to ask whether their emptiness problem is decidable,
and if it is, then in which complexity class it lies.

There are several results on decidability and computational complexity
of the emptiness problem
for simpler kinds of finite automata
that traverse an input object:
for deterministic two-way finite automata (2DFA),
the emptiness problem is PSPACE-complete
(this follows from the work of Kozen~\cite[Lemma~3.2.3]{Kozen}),
whereas for deterministic tree-walking automata
the analogous problem is EXP-complete, as proved by Boja\'nczyk~\cite{Bojanczyk}.

Another kind of finite automata
are nondeterministic automata
that recognize a given object by \emph{tiling} it with neighbourhoods of states.
Such are one-way nondeterministic finite automata (NFA),
for which the non-emptiness problem is NL-complete
(this is one of the classical problems presented by Jones~\cite{Jones}).
For trees, such are nondeterministic tree automata,
whose emptiness problem is P-complete, as shown by Veanes~\cite{Veanes}.

Tiling models were also considered for graphs.
Thomas~\cite{Thomas_tilings} introduced \emph{graph acceptors}:
in this model, a graph is accepted,
if it can be covered with tiles (subgraphs) from a fixed finite set,
so that each node is in the inner part of some tile,
states in overlapping tiles are the same,
and some further constraints on the number of occurrences of every tile hold.
For this general model, Thomas proved undecidability of the emptiness problem
by recognizing the set of rectangular grids and simulating a Turing machine on the grids.
Thomas also considered \emph{elementary acceptors}:
a special case in which every tile is a star, that is, a node with all its neighbours.
For elementary acceptors, Thomas proved that the language of grids cannot be recognized.
However, the decidability of the emptiness problem for elementary acceptors
remains open.

Besides the emptiness problem for graph-walking automata,
another problem considered in this paper
is the emptiness problem for \emph{star automata},
that is, for elementary acceptors of Thomas
without additional constraints on the number of occurrences of tiles.
Star automata are at the same time a special case of the model by Thomas,
and a generalization of nondeterministic tree automata to graphs.

In this paper, it is proved that the non-emptiness problem for graph-walking automata
is decidable, and furthermore, NEXP-complete,
while for star automata this problem is decidable and NP-complete.

The basic definitions of automata are given in Section~\ref{section_definitions}.
Graph-walking automata and star automata
are defined over a \emph{signature},
which is an alphabet for graphs.
A signature defines finite sets of possible node labels
and possible labels of edge end-points (called directions).
Also, for each node label, there is a set of directions
used in all nodes with this label.

The decidability of the emptiness problem
and upper bounds on its complexity
are obtained for graph-walking automata and for star automata
using similar methods.
A simpler problem called \emph{signature non-emptiness} is considered first:
does there exist at least one graph over a given signature?
Its decidability is proved in Section~\ref{section_emptiness_signature_NP}
by reducing it
to finding a non-negative integer solution to a certain system of linear equations.
From this, it is inferred that
the non-emptiness problem for signatures can be solved in NP.
Furthermore, if a signature is non-empty,
that is, if there is at least one graph over this signature,
then the number of nodes in the smallest such graph
does not exceed $2mr\min\{r^r,k^{2r-2}\}$,
where $m$ is the number of node labels in the signature,
$2r$ is the number of directions,
and $k$ is the maximum degree of a node.

It turns out that both checking non-emptiness of a graph-walking automaton
and checking non-emptiness of a star automaton
can be reduced to checking non-emptiness of a certain signature,
which is constructed for a given automaton.

For star automata, such a reduction
is presented in Section~\ref{section_star_to_signature}.
It gives a proof that the non-emptiness problem for star automata is in NP.
Also it gives an upper bound $sn^2k^{kn^2-1}$ on the number of nodes
in the smallest accepted graph,
where $n$ is the number of states in the star automaton,
$s$ is the number of stars,
and $k$ is the number of directions in the signature.

In Section~\ref{emptiness_gwa_to_signature},
a graph-walking automaton is reduced to a signature.
The reduction proves that its non-emptiness problem is in NEXP,
as well as gives an upper bound $m4^{n(k+1)}k^{k4^n-1}$
on the number of nodes in the smallest accepted graph,
where $n$ is the number of states,
$k$ is the number of directions,
and $m$ is the number of node labels.

In Section~\ref{NP_completeness_for_signature_and_NEXP_for_gwa},
all the above non-emptiness problems are proved to be hard in their complexity classes.
NP-hardness of the signature non-emptiness problem is obtained
by reducing 3-colourability to this problem.
This also gives NP-hardness for non-emptiness of star automata.
To prove NEXP-hardness of non-emptiness of graph-walking automata,
it is shown that a graph-walking automaton
can recognize the set of graphs containing a rectangular grid
of exponential size in the number of its states.
On this grid, the computation of a nondeterministic Turing machine
is then simulated.

Note that the complexity classes for related problems,
such as whether a graph-walking automaton accepts all graphs over its signature (the universality problem),
or whether the intersection of languages of two automata is empty,
can be inferred from the result for the non-emptiness problem.
Indeed, since every graph-walking automaton can be transformed to an automaton that halts on every input,
and the transformation given by Kunc and Okhotin~\cite{KuncOkhotin_reversible} can be done in polynomial time,
the emptiness problem for graph-walking automata
is equivalent to the universality problem.
As for the intersection emptiness problem,
Martynova and Okhotin~\cite{MartynovaOkhotin_gwa_boolean}
obtained a transformation for the intersection of two graph-walking automata,
which can be done in polynomial time too.
Thus, the universality problem and the intersection emptiness problem for graph-walking automata
are both co-NEXP-complete.

\section{Graph-walking and star automata}\label{section_definitions}

In this section, graph-walking automata and star automata are formally defined.
All definitions for graph-walking automata
are inherited from the paper by Kunc and Okhotin~\cite{KuncOkhotin_reversible}.
Star automata are a variant of \emph{elementary acceptors} by Thomas~\cite{Thomas_tilings}
without constraints on the number of tiles,
and are given in a different notation for uniformity with graph-walking automata.

Graph-walking automata are defined over a signature.
A signature specifies the sets of labels of nodes and edge end-points in the graphs,
and thus defines the set of all labelled graphs
that can be used as inputs for a graph-walking automaton.

\begin{definition}[\cite{KuncOkhotin_reversible}]
A \emph{signature} $S$ is a quintuple $S = (D, -, \Sigma, \Sigma_0, (D_a)_{a \in \Sigma})$,
where:
\begin{itemize}
\item
	$D$ is a finite set of directions,
	which are labels attached to edge end-points;
\item
	a bijection $- \colon D \to D$
	provides an opposite direction,
	with $-(-d) = d$ for all $d \in D$;
\item
	$\Sigma$ is a finite set of node labels;
\item
	$\Sigma_0 \subseteq \Sigma$
	is a subset of possible labels of the initial node;
\item
	$D_a \subseteq D$, for every
	$a \in \Sigma$,
	is the set of directions used in nodes labelled with $a$.
\end{itemize}
\end{definition}

Graphs are defined over a signature
like strings are defined over an alphabet.

\begin{definition}
A \emph{graph} over a signature $S = (D, -, \Sigma, \Sigma_0, (D_a)_{a \in \Sigma})$
is a quadruple $(V, v_0, +, \lambda)$, where:
\begin{itemize}
\item
	$V$ is a finite set of nodes;
\item
	$v_0 \in V$ is the initial node;
\item
	edges are defined by a partial function $+ \colon V \times D \to V$, such that
	if $v+d$ is defined, then $(v+d) + (-d)$ is defined and equals $v$;
	also denote $v-d = v+(-d)$;
\item
	node labels are assigned by
	a total mapping $\lambda \colon V \to \Sigma$, such that
	\begin{enumerate}\renewcommand{\theenumi}{\roman{enumi}}
	\item
	$v+d$ is defined
	if and only if
	$d \in {D_{\lambda(v)}}$,
	and
	\item
	$\lambda(v) \in \Sigma_0$
	if and only if
	$v = v_0$.
	\end{enumerate}
\end{itemize}
The set of all graphs over the signature $S$ is denoted by $L(S)$.
\end{definition}

The function $+$ defines the edges of the graph.
If $u+d = v$, then the nodes $u$ and $v$ in the graph
are connected with an edge with its end-points
labelled with directions $d$ (on the side of $u$)
and $-d$ (on the side of $v$).
Multiple edges and loops are possible:
if $v+d = v$ and $d \neq -d$, then it is a loop at the node $v$
with two ends labelled with directions $d$ and $-d$.
If $v+d = v$ and $d = -d$, then it is a loop at the node $v$ with one end,
labelled with $d$.

A graph-walking automaton is defined similarly to a 2DFA,
with an input graph instead of an input string.

\begin{definition}
\emph{A (deterministic) graph-walking automaton (GWA) over a signature 
$S = (D, -, \Sigma, \Sigma_0, (D_a)_{a \in \Sigma})$} 
is a quadruple $A = (Q, q_0, F, \delta)$, where
\begin{itemize}
\item
	$Q$ is a finite set of states;
\item
	$q_0 \in Q$ is the initial state;
\item
	$F \subseteq Q \times \Sigma$ is a set of acceptance conditions;
\item
	$\delta \colon (Q \times \Sigma) \setminus F \to Q \times D$ is 
	a partial transition function, with $\delta(q, a) \in Q \times D_a$ for all $q$ and $a$
	where $\delta$ is defined.
\end{itemize}
\end{definition}

When an automaton operates on a graph,
at every moment it knows its current state and sees only the label of the current node.
The transition function gives the new state and the direction to one of the neighbouring nodes,
in which the automaton moves.
If the current pair of a state and a node label is in $F$,
then the automaton accepts.
If the pair is not in $F$ and no transition is defined for it, then the automaton rejects.
It may also continue walking indefinitely, it this case it is said to loop.

Formally, an automaton's \emph{configuration} on a graph $G=(V, v_0, +, \lambda)$
is a pair $(q,v)$, with $q \in Q$ and $v \in V$. 
A \emph{computation} of an automaton $A$ on a graph $G$
is the following uniquely defined sequence of configurations.
The computation starts in the initial configuration $(q_0,v_0)$.
For every configuration $(q,v)$ in the computation,
if $\delta(q, \lambda(v))$ is defined and equals $(q', d)$,
then the next configuration after $(q,v)$ is $(q', v+d)$.
Otherwise, the configuration $(q,v)$ is the last one in the computation;
if $(q, \lambda(v)) \in F$, then the automaton \emph{accepts} in the configuration $(q,v)$,
otherwise it rejects.
If the computation is an infinite sequence, then the automaton is said to \emph{loop}.

A graph-walking automaton $A$ defines the language $L(A)$,
this is the set of graphs it accepts.

The methods used in this paper to prove the decidability of the emptiness problem
for graph-walking automata 
and to determine its computational complexity
can also be applied to another related model.
These are \emph{star automata}, which are defined as follows.

\begin{definition}
Let $S = (D, -, \Sigma, \Sigma_0, (D_a)_{a \in \Sigma})$ be a signature 
and let some linear order be fixed on the set of directions $D$.
\emph{A star automaton $A_*$ over the signature $S$} is a pair $(Q, T)$, where
\begin{itemize}
\item 
	$Q$ is a finite set of states;
\item
	$T$ is a finite set of stars, where
	a star is a sequence of the form $(a,q,q_1,\ldots,q_{|D_a|})$,
	where $a$ is a node label, $q$ is used for the state in the current node,
	$q_1, \ldots, q_{D_a}$ are used for states in the neighbours of the current node
	in all directions from $D_a$. 
\end{itemize}
\end{definition}

A graph $G$ is accepted by the star automaton $A_*$, if there is a choice of states $(q(v))_{v \in V}$
in all nodes such that the following condition holds for each node $v \in V$. 
Let $a$ be the label of the node $v$, let $d_1, \ldots, d_{D_a}$
be the directions from $D_a$ listed in the order. Then, the star in the node $v$
is the sequence $s(v) = (a,q(v),q(v+d_1), \ldots, q(v+d_{|D_a|}))$.
And every such star should belong to the set of automaton's stars $T$.
Such a sequence $(q(v))_{v \in V}$
is called \emph{a computation} of the star automaton $A_*$ on the graph $G$.
There can be several computations.

\section{The non-emptiness problem for signatures is in NP} \label{section_emptiness_signature_NP}

In this section, the decidability of the non-emptiness problem for signatures
is proved; more precisely, an NP-algorithm that solves this problem is constructed.
Furthermore, for non-empty signatures, an upper bound on the number of nodes 
in the minimal graph over a given signature is obtained.

It turns out that to prove that a signature is non-empty it is not necessary to find an actual graph.
It is sufficient to find only a collection of nodes without the edge structure of the graph;
such a collection is described by a vector with every coordinate giving the number of nodes with a certain label.
A vector can be turned into a graph if it satisfies a few conditions.

\begin{definition}
Let $S = (D, -, \Sigma, \Sigma_0, (D_a)_{a \in \Sigma})$ be a signature.
A vector of non-negative integers $(x_a)_{a \in \Sigma}$, where $x_a$ is 
the number of nodes with the label $a$,
is called \emph{balanced}, if it satisfies the following two balance conditions:
\begin{enumerate}
\item
	an initial node exists and is unique: $\sum_{a_0 \in \Sigma_0}x_{a_0} = 1$,
\item
	for each direction $d \in D$,
	such that $d \neq -d$,
	all nodes together need the same number of edges by $d$
	and by $-d$:
	\begin{equation*}
	\sum_{a \in \Sigma :\; d \in D_a}x_a = \sum_{a \in \Sigma:\;-d \in D_a}x_a.
	\end{equation*}
\end{enumerate}
\end{definition}

The next lemma shows that every balanced vector gives rise to a graph,
and hence one can work with balanced vectors instead of graphs.

\begin{lemma}\label{lemma_graphs_to_x_a}
Let $S = (D, -, \Sigma, \Sigma_0, (D_a)_{a \in \Sigma})$ be a signature.
Let $x_a$, for each node label $a \in \Sigma$,
be a non-negative integer.

A graph over the signature $S$ with exactly $x_a$ nodes labelled with $a$,
for all $a \in \Sigma$, exists if and only if the vector $(x_a)_{a \in \Sigma}$ is balanced.

Furthermore, there is an algorithm that, given a signature $S$
and a balanced vector $(x_a)_{a \in \Sigma}$,
constructs a graph over $S$ with exactly $x_a$ nodes with each label $a \in \Sigma$, 
and does so in time linear in the sum of
sizes of the signature and of the constructed graph.
\end{lemma}

\begin{proof}
For every graph $G$ over $S$, let $(x_a)_{a \in \Sigma}$ be the vector of quantities 
of nodes for all labels. It is claimed that
the vector $(x_a)_{a \in \Sigma}$ is balanced.
The first balance condition holds, because every graph has exactly one initial node.
Now to the second condition. Let $d \in D$ be one of the directions, with $d \neq -d$. 
Then, every edge $v+d = u$ in the graph links the two edge end-points: 
in the direction $d$ at the node $v$,
and in the direction $-d$ at the node $u$. Thus, 
the total number $\sum_{a \in \Sigma :\; d \in D_a}x_a$ of edge end-points
labelled with $d$ in the graph equals the number $\sum_{a \in \Sigma :\; -d \in D_a}x_a$
of edge end-points labelled with $-d$, and the second balance condition holds.

Conversely, let $(x_a)_{a \in \Sigma}$ be a balanced vector.
A graph $G = (V,v_0,+,\lambda)$ 
with exactly $x_a$ nodes for each node label $a$ 
is constructed by the following algorithm.

\begin{itemize}
\item
	First, the set of nodes $V$ and the labelling function $\lambda$ are defined: 
	for each node label $a \in \Sigma$ in the signature,
	$x_a$ new nodes
	labelled with $a$ are added to the set $V$.
\item
	The initial node is the node with a label from the set $\Sigma_0$,
	the first balance condition states that such a node exists and is unique.
\item
	Now the edges shall be defined so, that each node $v$ labelled with $a$
	will have edges exactly in the directions from $D_a$.
	For each direction $d \in D$, let $I_d$ be the set of all nodes $v$
	with $d \in D_{\lambda(v)}$.
	
	For such directions $d \in D$, that $d = -d$, the algorithm makes loops:
	for every node $v \in I_d$ it adds a loop $v+d = v$.
	
	For each pair of opposite directions $d \neq -d$,
	the algorithm takes nodes from $I_d$ and $I_{-d}$,
	and links them with $(d,-d)$-edges.
	By the second balance condition, $|I_d| = |I_{-d}|$,
	thus, every node gets all the edges it needs.
\end{itemize}
\end{proof}

Now, to check whether a signature is non-empty, that is, whether there is at least one graph
over this signature, one can just check whether there is at least one balanced vector for
this signature.

For a signature $S = (D, -, \Sigma, \Sigma_0, (D_a)_{a \in \Sigma})$,
balanced vectors $(x_a)_{a \in \Sigma}$ with the minimal possible sum
of coordinates will be called \emph{minimal balanced vectors}. 

How large could be the sum of the coordinates of a minimal balanced vector?
The next theorem gives an upper bound on this sum, that is, 
on the minimal number of nodes in the graph over a signature.

\begin{theorem} \label{theorem_small_graphs_signature}
Let $S = (D, -, \Sigma, \Sigma_0, (D_a)_{a \in \Sigma})$ be a non-empty signature,
and assume that $|D| \geqslant 2$
and that $D_a$ is non-empty for all $a \in \Sigma$.
Let $r = \frac{1}{2}|D|$,
$m=|\Sigma|$ and
$k = \max_{a \in \Sigma}|D_a|$.

Then, there is a graph over the signature $S$
with at most $2mr\min\{r^r, k^{2r-2}\}$ nodes.
\end{theorem}

Note that the bound $2mrk^{2r-2}$ can be useful for signatures with many directions,
but with a small maximum degree of nodes. Later on, such signatures
will be produced by the reductions of the emptiness problems for
graph-walking automata and for star automata to the emptiness problem for signatures.

First, the conditions and the claims of Theorem~\ref{theorem_small_graphs_signature}
are reformulated in the language of linear algebra.

By Lemma~\ref{lemma_graphs_to_x_a}, 
to prove Theorem~\ref{theorem_small_graphs_signature}
it is sufficient to prove that there is such a balanced vector $(x_a)_{a \in \Sigma}$
that $\sum_{a \in \Sigma} x_a \leqslant 2mr\min\{r^r, k^{2r-2}\}$.

Let $n$ be the number of pairs of opposite directions $\{d,-d\}$, with $d,-d \in D$
and $d \neq -d$, in the signature $S$.
It is convenient to rewrite $n$ linear equations in the second balance condition
as one vector equation. 
Let $\{d_1,-d_1, \ldots, d_n, -d_n\}$ be all such directions in $D$ that $d \neq -d$, 
here the directions $d_i$ and $-d_i$ are opposite, for $i = 1, \ldots, n$.

For each node label $a \in \Sigma$, the contribution of one node labelled with $a$ to 
the balance of directions in a graph is given by a column vector $v_a$
of height $n$. The $i$-th element of the vector $v_a$, for $i \in \{1, \ldots, n\}$,
is defined as follows:
\begin{equation}\label{equation_va}
	v_{a,i} = 
 	\begin{cases}
   		1 & \text{if } d_i \in D_a, -d_i \notin D_a\\
   		-1 & \text{if } -d_i \in D_a, d_i \notin D_a\\
   		0 & \text{if } d_i \notin D_a, -d_i \notin D_a \text{ or } d_i \in D_a, -d_i \in D_a
 	\end{cases}
\end{equation}
Thus, the $i$-th element of the vector $v_a$ is the contribution of an $a$-labelled node
to the difference of the number of directions $d_i$ and $-d_i$ in a graph.

Then, the second balance condition for the vector of quantities of labels $(x_a)_{a \in \Sigma}$
can be written in the following form:
\begin{equation*}
\sum_{a \in \Sigma}x_av_a = 0
\end{equation*}

If $n = 0$, then all directions are of the form $d = -d$, and one initial node with the loops
is a correct graph. Let $n \geqslant 1$.
As $n \leqslant r$, it is sufficient to prove an upper bound $2mn\min\{n^n, k^{2n-2}\}$.
Let $(x_a)_{a \in \Sigma}$ be a balanced vector with the minimal possible sum of the coordinates.
Among the initial node labels, only one has a non-zero coefficient.
Fix this initial label $a_0$ and let the vector $-v_{a_0}$ be denoted by $b$.
Then, the coefficients for other initial labels are zeros
and $\sum_{a \in (\Sigma \setminus \Sigma_0)}x_a
= \big(\sum_{a \in \Sigma}x_a\big) - 1$.
Then, to prove the theorem,
it is sufficient to find such a non-negative integer solution $(x_a)_{a \in (\Sigma \setminus \Sigma_0)}$
to the equation $\sum_{a \in (\Sigma \setminus \Sigma_0)}x_av_a = b$,
that $\sum_{a \in (\Sigma \setminus \Sigma_0)}x_a \leqslant
2mn\min\{n^n, k^{2n-2}\}-1$.

Some vectors $v_a$ for different non-initial labels can coincide.
Let $v_1, \ldots, v_\ell$ be all vectors from the set 
$\set{v_a}{a \in (\Sigma \setminus \Sigma_0)}$ without repetitions and without a zero vector. 
Note that $\ell < m$.
Then, it is sufficient to find a non-negative integer solution $(x_i)_{i = 1}^\ell$
to the equation $\sum_{i = 1}^\ell x_iv_i = b$, 
with $\sum_{i = 1}^\ell x_i \leqslant 2 \ell n\min\{n^n, k^{2n-2}\}$.

What is known about vectors $v_1, \ldots, v_\ell$?
These are column vectors of height $n$, 
with all elements in $\{0,1,-1\}$. 
Each vector has at most $k$ non-zero elements,
since each node label $a \in \Sigma$ has at most $k$ directions in $D_a$. 
To apply the methods of linear algebra, these vectors are considered over
the field of real numbers: $v_1, \ldots, v_\ell \in \mathbb{R}^n$.
Therefore, Theorem~\ref{theorem_small_graphs_signature} is reduced to the following lemma.

\begin{lemma} \label{lemma_linear_equations}
Let $v_1, \ldots, v_\ell \in \{0,1,-1\}^n$ be distinct non-zero column vectors of height $n$, 
where $n \geqslant 1$; let $b \in \{0,1,-1\}^n$ be a column vector. 
Let $k$ be the maximum number of non-zero elements in the vector.
Then, if the linear equation $\sum_{i = 1}^\ell x_iv_i = b$ has at least one non-negative integer
solution, then there exists such a non-negative integer solution $(x_i)_{i = 1}^\ell$,
that $\sum_{i = 1}^\ell x_i \leqslant 2\ell n\min\{n^n, k^{2n-2}\}$.
\end{lemma}

The proof of Lemma~\ref{lemma_linear_equations}
will use a classical bound on matrix determinants, as well as its corollaries.
Hadamard obtained the following upper bound for matrices of $-1$ and $1$.

\begin{oldtheorem}[Hadamard~\cite{Hadamard}]\label{theorem_Hadamard}
Let $n \geqslant 1$ be an integer and let $A$ be an $n \times n$ matrix,
with all elements in $\{-1,1\}$.
Then, $|{\det A}| \leqslant n^{\frac{n}{2}}$.
\end{oldtheorem}

Hadamard also proved that if $n$ is a power of $2$, then the bound $n^{\frac{n}{2}}$
is achieved on some matrices.

The upper bound by Hadamard can be generalized from the case of elements in $\{-1,1\}$ 
to any real numbers with absolute value not greater than $1$.
This is established in the following corollary. Also, I include a trivial upper bound
for matrices with a small number of non-zeros in columns.

\begin{corollary}\label{corollarynndet}
Let $n \geqslant 1$ be an integer and let $A$ be an $n \times n$ matrix, with all elements
real and not exceeding $1$ in absolute value.
Then, $|{\det A}| \leqslant n^{\frac{n}{2}}$.

If for some integer $k$, with $k \geqslant 1$, it is known that
each column in the matrix $A$, maybe except one,
has at most $k$ non-zero elements,
then $|{\det A}| \leqslant k^{n-1}$.
\end{corollary}

\begin{proof}
First, the upper bound $n^{\frac{n}{2}}$ is proved for matrices 
without conditions on the number of non-zero elements.
Among all $n \times n$ matrices with all elements real 
and not exceeding $1$ in absolute value,
let $A$ be the one with the maximum absolute value of the determinant.
It can be assumed that $\det A \geqslant 0$, 
because otherwise one can multiply the first column by $-1$ and negate the determinant.
The determinant of $A$ is given by the following formula.
\begin{equation*}
\det A = \sum_{\sigma \in S_n} (-1)^{\text{sign}(\sigma)}\prod_{i = 1}^n a_{i,\sigma(i)}.
\end{equation*}
If all elements of $A$ are in $\{-1,1\}$, then $|{\det A}| \leqslant n^{\frac{n}{2}}$
by the bound by Hadamard (Theorem~\ref{theorem_Hadamard}).
Now let $a_{ij}$ be any element in $A$ with absolute value less than $1$.
The formula for the determinant can be represented as ${\det A} = ba_{ij}+c$,
where $b$ and $c$ do not depend on $a_{ij}$. 
If $b$ is positive, then $a_{ij}$ can be changed to $1$,
making the determinant of $A$ greater; otherwise, $a_{ij}$ can be replaced with $-1$,
without decreasing the determinant.
Thus, all elements with absolute value less than $1$ can be replaced one by one
without decreasing the determinant, and so the Hadamard's bound
$|{\det A}| \leqslant n^{\frac{n}{2}}$ holds for the matrix $A$.

Now to the second part of the corollary. 
Let the matrix $A$ have at most $k$ non-zero elements
in each column, maybe except one column. By induction on $n$
it is proved that the determinant of every such matrix is at most $k^{n-1}$.
For a $1 \times 1$ matrix, the determinant does not exceed $1$ in absolute value.
For an $n \times n$ matrix, with $n > 1$,
as the absolute value of the determinant does not change 
when the columns in the matrix are permuted, 
one can assume that the last column has
the greatest number of non-zero elements.
By expanding along the first column, the determinant
equals a sum of not more than $k$ determinants of $(n-1) \times (n-1)$ 
matrices with the same properties, which by induction hypothesis are not greater
than $k^{n-2}$ in absolute value. 
And these matrices are taken with coefficients not greater than 1 in absolute value.
Thus, $|{\det A}| \leqslant k \cdot k^{n-2} =  k^{n-1}$.
\end{proof}

The upper bounds on determinants are used to estimate the coefficients
in linear equations.

\begin{lemma}\label{lemma_linear_combination}
Let $n \geqslant 1$ be an integer, let $v_1,\ldots, v_t \in \{0,1,-1\}^n$, with $t \geqslant 1$, 
be column vectors of height $n$, which are linearly independent in $\mathbb{R}^n$.
Let $k$ be the maximum number of non-zero elements in a vector,
and let $N = \min\{n^{\frac{n}{2}}, k^{n-1}\}$.
Let some vector $u \in \mathbb{R}^n$, with the maximum absolute
value of its elements $c$, be represented as a linear combination:
$\alpha_1v_1+\ldots+\alpha_tv_t = u$.

Then, $|\alpha_i| \leqslant cN, \text{ for all } i \in \{1, \ldots, t\}$.
Furthermore, if all elements in the vector $u$ are integers, then all coefficients $\alpha_i$, 
for $i \in \{1, \ldots, t\}$, are rational, and after multiplying the equation by
their least common denominator one obtains the equation
$\beta_1v_1+\ldots+\beta_tv_t+\beta_{t+1}u = 0$, with all coefficients $\beta_i$, 
for $i \in \{1, \ldots, t+1\}$, integer and not exceeding $cN$ in absolute value.
\end{lemma}

\begin{proof}
If $u$ is a zero vector, then all coefficients in the linear combination are zeros. 
Now let $u$ be not a zero vector.
The vectors $v_1, \ldots, v_t$ are linearly independent, so 
the system of equations $x_1v_1+\ldots+x_tv_t = u$ has at most one solution.
Thus, the solution $(\alpha_1, \ldots, \alpha_t)$ is unique.
To solve this system of equations using Cramer's rule, one needs the matrix
of coefficients $V = (v_1,\ldots,v_t)$ to be square.

Since the vectors $v_1, \ldots, v_t$ are linearly independent, $t \leqslant n$. 
First, consider the case of $t < n$.
The matrix $(v_1,\ldots,v_t,u)$ has the column rank $t$, because
the columns $v_1, \ldots, v_t$ are linearly independent, and the column $u$ 
is their linear combination. 
It is known that the column rank equals the row rank,
so there are $t$ linearly independent rows in the matrix $(V,u)$,
all other rows are their linear combinations. That is, 
in the system of equations $x_1v_1+\ldots+x_tv_t = u$,
all equations are linear combinations of some $t$ linearly independent equations.
Taking only these $t$ linearly independent equations
one obtains a system $x_1v_1'+ \ldots+x_tv_t' = u'$, with all vectors
of height $t$. The set of solutions has not changed,
so $(\alpha_1, \ldots, \alpha_t)$ remains the only solution.
Let $V' = (v_1', \ldots, v_t')$ be the matrix of coefficients 
of the new system of equations,
it is a non-degenerate square matrix.
If $t = n$, then the matrix $V$ is already square and non-degenerate;
in this case let $V' = V$, $u' = u$.

Now the new system of equations can be solved by Cramer's rule.
Let $V_i' = (v_1',\ldots,v_{i-1}',u',v_{i+1}',\ldots,v_t')$ be the matrix,
obtained from $V'$ by replacing of the $i$-th column with the column vector $u'$,
for each $i = 1, \ldots, t$.
Then, Cramer's rule claims that
the unique solution to the system is $\alpha_i = \frac{\det V_i'}{\det V'}$, for $i = 1,\ldots, t$.

Now one needs to estimate the determinants of the matrices $V'$ and $V'_i$, for $i = 1, \ldots, t$.
The matrix $V'$ has all its elements in $\{0,1,-1\}$.
Also, each column of $V'$ has at most $k$ non-zeros.
So Corollary~\ref{corollarynndet} gives
$|{\det V'}| \leqslant \min\{n^{\frac{n}{2}},k^{n-1}\} = N$.
Since all elements of $V'$ are integers and the matrix is non-degenerate,
${\det V'}$ is a non-zero integer.
Now consider the matrix $V_i'$, for some $i = 1, \ldots, t$.
Let $V_i''$ be the matrix obtained from $V_i'$ by dividing
the $i$-th column, which equals $u$, by $c$.
Then, all elements of $V_i''$ are not greater than $1$ in absolute value.
And each column has at most $k$ non-zero elements, maybe, except the $i$-th column.
By Corollary~\ref{corollarynndet}, the determinant of $V_i''$ is estimated
as follows: $|{\det V_i''}| \leqslant \min\{n^{\frac{n}{2}},k^{n-1}\} = N$. 
Thus, the determinant of the matrix $V_i'$, which has one column multiplied by $c$,
is bounded like this: $|{\det V_i'}| \leqslant cN$.

So, $|\alpha_i| = |\frac{\det V_i'}{\det V'}| \leqslant |{\det V_i'}| \leqslant cN$, 
for all $i = 1, \ldots, t$.
If all elements of the vector $u$ are integers, then all the determinants ${\det V_i'}$
are integers as well. Then all $\alpha_i$, for $i = 1, \ldots, t$, are rational.
And after muliplying the equation by their least common denominator, which is not greater
than $|{\det V'}|$ in absolute value,
one gets all new coefficients $\beta_i$, for $i = 1, \ldots, t+1$, 
not greater in absolute value than $\max \{|{\det V'_1}|, \ldots, |{\det V'_t}|, |{\det V'}|\} \leqslant cN$.
\end{proof}

Now it is time to prove the lemma, to which
Theorem~\ref{theorem_small_graphs_signature} has been reduced.

\begin{proof}[Proof of Lemma~\ref{lemma_linear_equations}]
Let $N = \min\{n^{\frac{n}{2}},k^{n-1}\}$ be the upper bound 
from Corollary~\ref{corollarynndet}
on the determinants of $n \times n$ matrices with real elements
not exceeding $1$ in absolute value, and with at most $k$ non-zero elements in each column,
maybe, except one.

Let $(x_1, \ldots, x_\ell)$ be a non-negative integer solution 
to the system of linear equations $\sum_{i = 1}^\ell x_iv_i = b$,
with the minimum sum $\sum_{i = 1}^\ell x_i$, and among these,
with the minimum number of coordinates greater than $N$.
The goal is to prove, that $\sum_{i = 1}^\ell x_i \leqslant 2\ell n\min\{n^n, k^{2n-2}\}$.

\textbf{Step 1} is to prove that all vectors $v_i$, for $i = 1, \ldots, \ell$, with $x_i > N$,
are linearly independent over the field $\mathbb{R}$.

For the sake of a contradiction, suppose that these vectors are linearly dependent.
Then a linear dependence involving the least number of vectors is chosen.
The vectors $v_1, \ldots, v_\ell$ are rearranged, so that
the vectors from the dependence go in the beginning:
let $v_1, \ldots, v_{t+1}$ be the vectors from this minimal linear dependence.
It is known that $t \geqslant 2$, because all vectors $v_1, \ldots, v_\ell$ are distinct and 
there is no zero vector among them. 

The vectors $v_1, \ldots, v_t$ are linearly independent, whereas $v_1, \ldots, v_{t+1}$
are linearly dependent. Then, the vector $v_{t+1}$ is uniquely represented 
as a linear combination of the others: 
$v_{t+1} = \alpha_1v_1+\ldots+\alpha_tv_t$, 
where $\alpha_1,\ldots, \alpha_t \in \mathbb{R}$.

The vector $v_{t+1}$ has all its elements integer and the maximum absolute value of its elements is $1$; the vectors $v_1, \ldots, v_t$ satisfy all conditions of Lemma~\ref{lemma_linear_combination}.
Thus, by Lemma~\ref{lemma_linear_combination}, all coefficients $\alpha_1,\ldots,\alpha_t$ are rational, and after multiplying
the linear combination by their least common denominator
one gets the new linear combination $\beta_1v_1+\ldots+\beta_{t+1}v_{t+1} = 0$,
with all coefficients integer and not exceeding $N$ in absolute value.

Since the chosen linear dependence has the minimal number of vectors,
$\beta_i \neq 0$, for all $i = 1, \ldots, t+1$.
If $\sum_{i =1}^{t+1} \beta_i < 0$,
then the dependence $\beta_1v_1+\ldots+\beta_{t+1}v_{t+1} = 0$ can be multiplied by $-1$,
so one can assume, that $\sum_{i =1}^{t+1} \beta_i \geqslant 0$.

Consider the case when $\sum_{i =1}^{t+1} \beta_i > 0$. 
Then, let $(y_1, \ldots, y_\ell)$ be a vector defined 
by $y_i = x_i-\beta_i$, for $i = 1, \ldots, t+1$,
and $y_i = x_i$, for $i = t+2, \ldots, \ell$.
Then, $\sum_{i = 1}^\ell y_iv_i = b$, that is, $(y_1, \ldots, y_\ell)$ is another solution 
to the system of equations. All $y_i$ are non-negative integers,
because $x_1, \ldots, x_{t+1}$ are greater than $N$, and $\beta_1, \ldots, \beta_{t+1}$
are integer and not greater than $N$ in absolute value. 
And, $\sum_{i = 1}^\ell y_i < \sum_{i = 1}^\ell x_i$.
This contradicts the minimality
of the sum of the coordinates in the solution $(x_1, \ldots, x_\ell)$.

Now let $\sum_{i =1}^{t+1} \beta_i = 0$.
Then one can similarly subtract $(\beta_1,\ldots,\beta_{t+1})$ from $(x_1, \ldots, x_{t+1})$ 
several times until some coefficient among the first $t+1$
becomes not greater than $N$. Such subtractions will not break the equation,
will not make any coordinate negative,
will not change the sum of the coordinates in the solution,
but will decrease the number of coordinates which are greater than $N$. 
This contradicts the minimality of the number of such coordinates among 
the solutions with the minimal sum of the coordinates.

Step 1 is done. Now it is known that all vectors among $v_1, \ldots, v_\ell$
which have the corresponding coefficients in the solution $(x_1, \ldots, x_\ell)$ 
greater than $N$ are linearly independent. Let these vectors be put first, 
so that they are $v_1, \ldots, v_t$.

\textbf{Step 2} is to prove that $x$ is the desired solution, that is, that
$\sum_{i = 1}^\ell x_i \leqslant 2\ell n\min\{n^n, k^{2n-2}\}$.

The sum to be estimated is: 
$\sum_{i = 1}^\ell x_i = \sum_{i=1}^t x_i+\sum_{i = t+1}^\ell x_i$. 
The second sum is bounded by 
$\sum_{i = t+1}^\ell x_i \leqslant (\ell-t)N$, as it has all coefficients not greater than $N$.
If the first sum is non-empty ($t > 0$), then 
the first $t$ variables are bounded as follows.
The system of equations is rewritten in the following way: 
$x_1v_1 + \ldots + x_tv_t = b-(x_{t+1}v_{t+1}+\ldots+ x_\ell v_\ell)$.
Here the vectors $v_1, \ldots, v_t$ are linearly independent, whereas the sum
on the right-hand side is a column vector of height $n$, with all elements
not greater than $\ell N$ in absolute value (if $t > 0$, then $\sum_{i = t+1}^\ell x_i < \ell N$).
By applying Lemma~\ref{lemma_linear_combination}, 
with $u = b-(x_{t+1}v_{t+1}+\ldots+ x_\ell v_\ell)$,
one obtains $|x_i| \leqslant \ell N^2$, for all $i = 1, \ldots, t$.
As $t \leqslant n$,
\begin{multline*}
\sum_{i = 1}^\ell x_i = \sum_{i=1}^t x_i+\sum_{i = t+1}^\ell x_i
\leqslant t \ell N^2+(\ell-t)N \leqslant n \ell N^2+ \ell N \leqslant \\ 
\leqslant 2 \ell nN^2 = 2\ell n\min\{n^n,k^{2n-2}\}.
\end{multline*}
\end{proof}

Theorem~\ref{theorem_small_graphs_signature}, which has just been proved,
gives the upper bound $2mr\min\{r^r, k^{2r-2}\}$ on the number of
nodes in the minimal graph over a non-empty signature, which depends on its
parameters: on the number of node labels $m = |\Sigma|$,
on the number of directions $2r = |D|$ and
on the maximum possible degree of a node $k = \max\set{|D_a|}{a \in \Sigma}$.
This bound, and also Lemma~\ref{lemma_graphs_to_x_a},
that allows one to work with balanced vectors instead of graphs,
help to construct an NP-algorithm, that solves the non-emptiness problem
for signatures.

\begin{theorem}\label{theorem_signature_non_emptiness_is_in_NP}
There is an NP-algorithm that takes a signature as an input and
determines whether there is at least one graph over this signature or not.
\end{theorem}

\begin{proof}
The size of an input $S = (D, -, \Sigma, \Sigma_0, (D_a)_{a \in \Sigma})$
is not less than $|\Sigma|+|D|$.
In the degenerate case of $|D| \leqslant 1$, it is sufficient to check for one-node graphs.
Any initial labels $a_0$ with $D_{a_0}$ empty form correct graphs;
any such non-initial labels can be omitted.

With the trivial cases removed,
by Theorem~\ref{theorem_small_graphs_signature}, if a graph over the signature $S$ exists,
then there is a graph with at most exponentially many nodes in $|D|$ and $|\Sigma|$.
Then, by Theorem~\ref{theorem_small_graphs_signature} 
and by Lemma~\ref{lemma_graphs_to_x_a}, the signature is non-empty 
if and only if there exists a balanced vector $(x_a)_{a \in \Sigma}$,
with the sum of coordinates not greater than this exponential upper bound.

Thus, the nondeterministic algorithm guesses a vector $(x_a)_{a \in \Sigma}$,
with sum of the coordinates not greater than exponential,
and writes it down in polynomial time.
It remains to check whether the guessed vector is balanced:
that is, whether only one label among the initial node labels has a non-zero coefficient, 
and whether for each pair of opposite directions $(d,-d) \in D$, with $d \neq -d$,
the following equation holds:
\begin{equation*}
	\sum_{a \in \Sigma :\; d \in D_a}x_a = \sum_{a \in \Sigma:\;-d \in D_a}x_a.
\end{equation*} 
This can all be checked in polynomial time, because the number of terms in these sums 
is polynomial, and each term is not greater than exponential.

If the algorithm guessed the vector, which is balanced, then the signature is non-empty 
and the algorithm answers ``yes''. Otherwise, it answers ``no''.
\end{proof}

In fact, the non-emptiness problem for signatures is NP-complete,
this is shown later in Section~\ref{NP_completeness_for_signature_and_NEXP_for_gwa}.

\section{Reducing a star automaton to a signature}\label{section_star_to_signature}

This section proves the decidability of the emptiness problem for star automata.
An NP-algorithm is constructed, which, for a given star automaton,
determines whether it accepts at least one graph.
Moreover, an upper bound on the number of nodes in the smallest accepted
graph is proved in this section.

It turns out that the emptiness problem for star automata
can be reduced in polynomial time to the emptiness problem for signatures,
which was proved to be in NP.

\begin{theorem}\label{theorem_star_to_signature}
There exists a polynomial-time algorithm that takes as an input
a signature $S = (D, -, \Sigma, \Sigma_0, (D_a)_{a \in \Sigma})$
with $k$ directions
and a star automaton $A_* = (Q, T)$ over $S$
with $n$ states and $s$ stars, and computes a signature
$S' = (D', -, \Sigma', \Sigma_0', (D'_{a'})_{a' \in \Sigma'})$
with $kn^2$ directions and with $s$ node labels,
with the following property.
There exists a bijective function $f$ that maps a graph $G$ over $S$
and a computation $C = (q(v))_{v \in V}$ of the automaton $A_*$ on this graph
to a graph $G' = f(G,C)$ over the signature $S'$,
which 
has the same set of nodes and the same edge structure as the graph $G$
(the only difference between $G$ and $G'$ is
in node labels and in directions).
\end{theorem}

\begin{proof}
Node labels and directions of the new signature $S'$ will contain
information on old node labels and directions, and also
some additional information that encodes the computation of the star automaton $A_*$
on a graph. More precisely, node labels will additionally encode stars
in nodes that appear in the computation, whereas directions will encode 
the states of the star automaton at the two ends of an edge.

The new signature $S'$ is constructed as follows.
\begin{itemize}
\item
	Node labels are all the stars of the automaton $A_*$, that is, $\Sigma' = T$.
\item
	Initial node labels are all the stars of $A_*$,
	in which the first component is an initial node label from the old signature, that is,
	$\Sigma_0' = \set{(a,q,q_1, \ldots, q_{|D_a|}) \in T}{a \in \Sigma_0}$.
\item
	The set of directions is $D' = D \times Q \times Q$,
	where the direction $(d,q_1,q_2)$ means
	that in the old graph the direction $d$ was here,
	and in the encoded computation the state at the current node is $q_1$
	and the state at the opposite end of the edge is $q_2$.
\item
	The relation of the opposite direction is: $-(d,q_1,q_2) = (-d,q_2,q_1)$,
	for all $(d,q_1,q_2) \in D'$.
\item
	For each star $t =(a,q,q_1,\ldots,q_{|D_a|}) \in \Sigma'$,
	where $d_1, \ldots, d_{|D_a|}$ are ordered directions from $D_a$, 
	the set of directions for the node label $t$ is defined by 
	$D'_t =  \set{(d_i,q,q_i)}{i = 1, \ldots, |D_a|}$.
\end{itemize}
Such a signature $S'$ can be computed from $S$ and $A_*$ in polynomial time.
There are exactly $kn^2$ directions and exactly $s$ node labels in the signature $S'$.

\begin{figure}[t]
	\centerline{\includegraphics[scale=0.9]{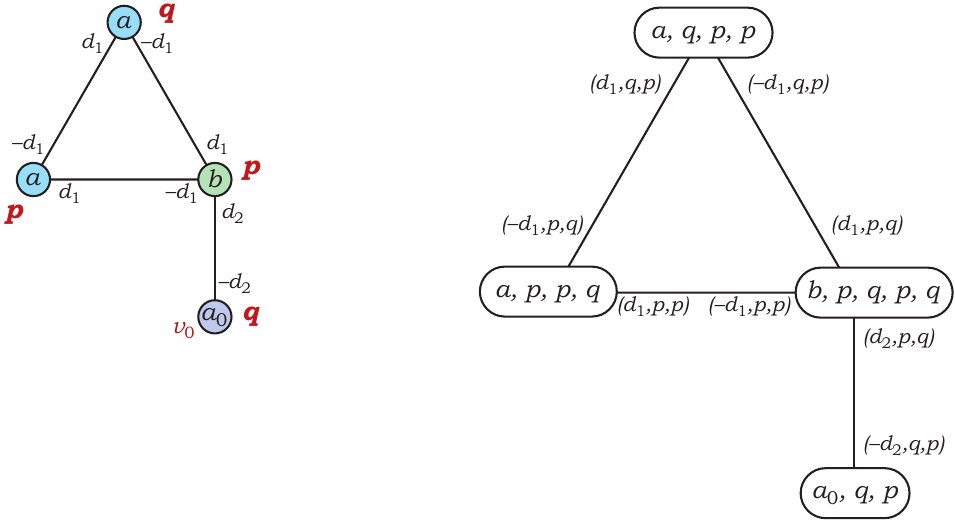}}
	\caption{Left: computation of a star automaton $A_*$ on a graph $G$. 
	Right: augmented graph $G'$ that encodes both $G$ and this computation.}
	\label{f:star_to_signature_h}
\end{figure}

It will be proved now, that there is a one-to-one correspondence
between graphs over $S'$
and pairs $(G, C)$ of a graph over $S$
and a computation of $A_*$ on this graph.
An example of such a correspondence is shown in Figure~\ref{f:star_to_signature_h}. 
For a star automaton $A_*$ 
with stars $(a,q,p,p)$, $(a,p,p,q)$, $(b,p,q,p,q)$, $(a_0,q,p)$,
its computation on a graph $G$ is given on the left.
On the right, there is a graph $G'$ that encodes stars in node labels and
states at the two ends of an edge in directions.

Let $G = (V,v_0,+,\lambda)$ be a graph over $S$, 
and let $C = (q(v))_{v \in V}$ be a computation of the star automaton $A_*$ on this graph.
Then the graph $f(G,C) = G' = (V',v_0',+,\lambda')$ over the signature $S'$
that encodes the graph $G$ and the computation $C$
is constructed as follows.
\begin{itemize}
\item
	The set of nodes and the initial node are the same: $V' = V$, $v_0' = v_0$. 
\item
	The edges in the graph $G'$ connect the same nodes as in $G$,
	but all the directions are augmented with the states at the ends of an edge.
	If $v+d = u$ in the graph $G$, then $v+(d,q(v),q(u)) = u$ in the graph $G'$,
	and these are all edges in $G'$.
	Then, the ends of each edge are labelled with opposite directions.
\item
	The node labels in $G'$ are stars in nodes.
	For each node $v \in V$ with some label $\lambda(v) = a$,
	the node label in the graph $G'$ is 
	$\lambda'(v) = (a,q(v),q(v+d_1), \ldots, q(v+d_{|D_a|}))$,
	where $d_1, \ldots, d_{|D_a|}$ are ordered directions from $D_a$.
	Then, $\lambda'(v) \in T = \Sigma'$, because $(q(v))_{v \in V}$ is a computation.
	And the directions in $G'$, used at the node $v$, are all the directions from $D_{\lambda'(v)}$.
	And only the initial node has an initial label.
\end{itemize}

This transformation maps different pairs $(G, C)$ to different graphs $G'$,
because no information is lost.
Conversely, for each graph $G'$ over the signature $S'$
there is a unique corresponding pre-image $(G, C)$,
where $G$ is obtained by dropping some information from all labels,
and node labels explicitly give states and stars in a computation.
Each edge in $G'$ checks that the states at the nodes it connects
are consistent with the stars.
\end{proof}

Now the results proved for signatures in the previous section 
will be transferred to star automata.

\begin{corollary}\label{theorem_nonemptiness_star_is_in_NP}
The non-emptiness problem for star automata, that is,
whether a given star automaton accepts at least one graph or not,
can be solved in NP.
\end{corollary}
\begin{proof}
By Theorem~\ref{theorem_star_to_signature},
for a star automaton $A_*$ that works over some signature $S$,
one can construct in polynomial time such a signature $S'$ of polynomial size,
that graphs over $S'$ are bijectively mapped
to the computations of $A_*$ on graphs over $S$.

A graph is accepted by the star automaton $A_*$
if there exists at least one computation of $A_*$ on it.
Thus, to check whether the star automaton is non-empty,
one can just check whether the signature $S'$ is non-empty.
By Theorem~\ref{theorem_signature_non_emptiness_is_in_NP},
the latter can be done in nondeterministic polynomial time.
\end{proof}

The upper bound on the number of nodes in the minimal graph over a signature
(Theorem~\ref{theorem_small_graphs_signature})
can be transferred to star automata as well.

\begin{corollary}\label{corollary_small_graphs_star_automata}
Let $S = (D, -, \Sigma, \Sigma_0, (D_a)_{a \in \Sigma})$ be a signature
with $k \geqslant 2$ directions,
and with $|D_a| \geqslant 1$ for all $a \in \Sigma$.
Let $A_* = (Q, T)$ be a star automaton with $n$ states and with $s$ stars over this signature.
If $A_*$ accepts at least one graph, then
the accepted graph with the minimal number of nodes
has at most $sn^2k^{kn^2-1}$ nodes.
\end{corollary}
\begin{proof}
Let $A_*$ accept at least one graph. 
The signature $S'$ is constructed from the signature $S$ and from the star automaton $A_*$
by Theorem~\ref{theorem_star_to_signature}.
The graphs over $S'$ correspond to the computations of $A_*$
on graphs over $S$ with the same number of nodes.

Then, the number of nodes in the minimal accepted graph for $A_*$
equals the number of nodes in the minimal graph over the signature $S'$.
This signature has $kn^2$ directions and $s$ node labels,
the maximum degree of a node does not exceed $k$ (because the function $f$ from
Theorem~\ref{theorem_star_to_signature} does not change the edge structure of a graph).
Then, Theorem~\ref{theorem_small_graphs_signature}
gives the following upper bound on the number of nodes in the minimal graph: 
$2s\frac{1}{2}kn^2\min\{{(\frac{1}{2}kn^2)}^{\frac{1}{2}kn^2}, k^{kn^2-2}\}$.
It can be bounded by a simpler expression:
\begin{equation*}
2s\frac{1}{2}kn^2\min\Big\{{(\frac{1}{2}kn^2)}^{\frac{1}{2}kn^2}, k^{kn^2-2}\Big\} \leqslant
skn^2k^{kn^2-2} = sn^2k^{kn^2-1}.
\end{equation*}
\end{proof}

\section{Reducing a graph-walking automaton to a signature}\label{emptiness_gwa_to_signature}

In Section~\ref{section_star_to_signature}, the emptiness problem for star automata
was reduced to the emptiness problem for signatures.
In this section such a reduction is made for the emptiness problem for graph-walking automata.

Note that whereas a computation of a star automaton is a way to choose states in nodes,
and the graph is accepted by a star automaton if there is at least one computation on this graph,
graph-walking automata are different.
In a graph-walking automaton,
the computation on a graph is a sequence of configurations $(q,v)$
of the automaton on a graph, where $q$ is the current state,
and $v$ is the node which the automaton visits at the moment.
This sequence in defined uniquely for each graph. 
The graph is accepted if the computation is accepting, that is, ends with an accepting configuration.

One way to reduce a graph-walking automaton to a signature
is to simulate it by a star automaton.
The next theorem shows that if some set of graphs is recognized by a graph-walking automaton,
then this set of graphs can be defined by some star automaton.
There is an analogous result for trees:
star automata on trees are \emph{nondeterministic tree automata},
graph-walking automata on trees are \emph{deterministic tree-walking automata},
and, as noted by Boja\'nczyk and Colcombet~\cite{BojanczykColcombet_det},
the inclusion of the class of languages defined even by nondeterministic tree-walking automata
into the class defined by tree automata is a folklore result.

\begin{theorem}\label{theorem_star_automaton_is_stronger_than_gwa}
For every $n$-state graph-walking automaton $A = (Q, q_0, F, \delta)$
over some signature $S= (D, -, \Sigma, \Sigma_0, (D_a)_{a \in \Sigma})$ 
with $k$ directions and $m$ node labels,
there exists a star automaton $A_* = (P, T)$ 
with $(k+1)^n$ states and at most $m(k+1)^{n(k+1)}$ stars,
defined over the same signature $S$,
which accepts exactly the same graphs as $A$.
The star automaton $A_*$ has size exponential in the size of $A$ 
and is constructed in exponential time.
\end{theorem}

This theorem is given without a proof,
because the next theorem gives a direct reduction of a graph-walking automaton to a signature
that provides a better upper bound
on the number of nodes in the minimal accepted graph.

\begin{theorem} \label{theorem_gwa_to_signature}
There exists an algorithm that takes as an input
some $n$-state graph-walking automaton $A = (Q, q_0, F, \delta)$ over some
signature $S = (D, -, \Sigma, \Sigma_0, (D_a)_{a \in \Sigma})$
with $k$ directions and $m$ node labels,
and computes such a signature 
$S' = (D', -, \Sigma', \Sigma_0', (D'_{a'})_{a' \in \Sigma'})$
with $k4^n$ directions and with not more than $m4^{nk}$ node labels,
that the following condition holds.

There exist two functions $f$ and $g$.
The function $f \colon L(A) \to L(S')$ injectively maps graphs over $S$, 
accepted by the automaton $A$, to graphs over $S'$,
and the function $g \colon L(S') \to L(A)$ is a surjection, such that $g(f(G)) = G$.
If $G' = f(G)$ or $G = g(G')$, then the graphs $G$ and $G'$
have the same sets of nodes and the same edge structure,
only node labels and directions are different.

The size of the resulting signature is exponential in the size of the input,
and the algorithm works in time exponential in the size of the input.
\end{theorem}

\begin{proof}
New node labels and directions of the signature $S'$ encode
node labels and directions of the signature $S$ and some additional information
about the behavior of the automaton $A$ in the vicinity of the node or edge end-point.

The new directions are 
$D' = D \times 2^Q \times 2^Q = \set{(d,Q_{in},Q_{out})}{d \in D;\, Q_{in},Q_{out} \subseteq Q}$.
Every new direction $(d,Q_{in},Q_{out})$
is an old direction $d$ with two sets of states attached:
$Q_{in}$ encodes the states in which the automaton came in its computation on a graph 
to the current edge end-point moving in the direction $-d$, 
whereas $Q_{out}$ consists of states, in which the automaton comes to the opposite end
of the edge, moving in the direction $d$.

The opposite direction is $-(d,Q_{in},Q_{out}) = (-d,Q_{out},Q_{in})$, for each $(d,Q_{in},Q_{out}) \in D'$.

Each node label in $S'$ contains an old node label and all information
about the new directions in the node.
But not every combination of new directions at a node
makes a new label.
The goal is to ensure that each graph over $S'$
encodes a graph over $S$ that is accepted by $A$,
along with an accepting computation of $A$ on this graph.
For this, some combinations that cannot appear
in accepting computations of the automaton $A$ will be left out. 

The set of node labels $\Sigma'$ is a subset of
\begin{equation*}
\widehat{\Sigma'} = \set{(a,E)}{a \in \Sigma, \: E = \{(d,Q_{in,d},Q_{out,d})\}_{d \in D_a},
	\text{ where } Q_{in,d},Q_{out,d} \subseteq Q \text{ for all } d \in D_a}.
\end{equation*}
It will be specified later, which elements of the set $\widehat{\Sigma'}$ are in $\Sigma'$ 
and which are not.

The set of directions of a new node label $(a,E)$ is $D'_{(a,E)} = E$.
The label $(a,E)$ is initial if and only if the label $a$ is initial.
Note that for each node label $a \in \Sigma$,
there is only one direction  $(d,Q_{in,d},Q_{out,d})$ with the first component $d$ in the set $E$,
for each direction $d \in D_a$.

\begin{figure}[t]
	\centerline{\includegraphics[scale=0.9]{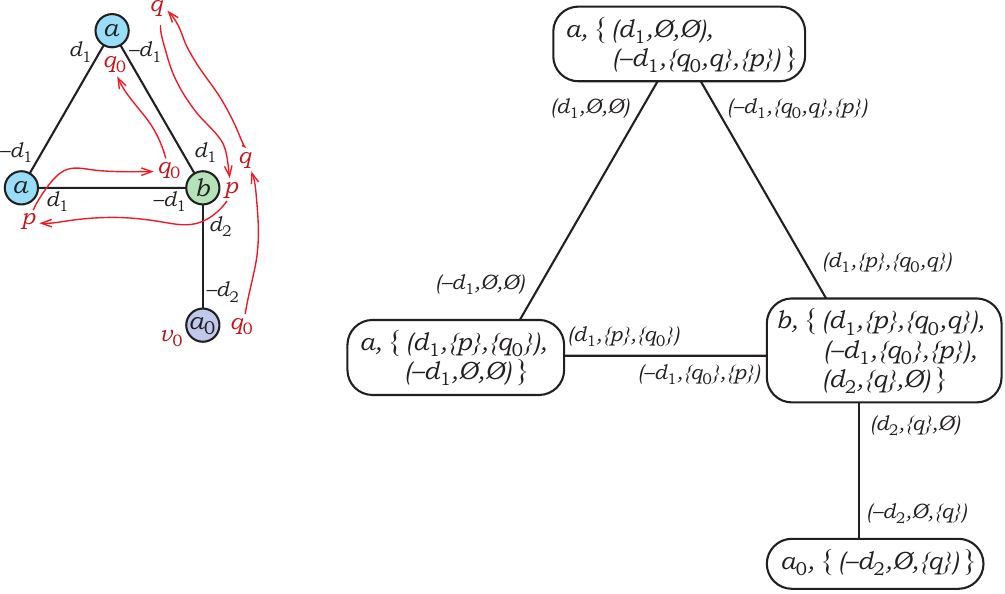}}
	\caption{Left: the accepting computation of some graph-walking automaton $A$
	on some graph $G$ over the signature $S$.
	Right: the graph $G'$ over the signature $S'$ 
	that encodes the graph and the accepting computation.}
	\label{f:gwa_to_signature_h}
\end{figure}

Figure~\ref{f:gwa_to_signature_h} gives an example of how
a graph $G$ over $S$ accepted by the automaton $A$ can be converted
to a graph $G'$ over the signature $S'$ 
by adding to each direction the information on the states 
in which the automaton crosses the edge, and by adding to each node label
the information contained in all new directions at the node.

To complete the definition of the signature $S'$, 
it remains to say, which pairs $(a,E)$ from the set
$\widehat{\Sigma'}$
are in the set $\Sigma'$, that is, are node labels of $S'$.
Some pairs $(a, E)$, which represent situations that cannot occur in any accepting computations of $A$,
will be left out,
and leaving them out will ensure
that every graph over the signature $S'$
encodes some graph $G$ and an accepting computation of $A$ on $G$.

A pair $(a,E)$ is in $\Sigma'$ if and only if the following conditions hold. 
\begin{enumerate}
\item
	The sets $Q_{in,d}$ and $Q_{in,e}$
	cannot intersect for directions $d \neq e$, where $d,e \in D_a$.
	If the label $a$ is initial, then for each $d \in D_a$ it is prohibited to have $q_0 \in Q_{in,d}$.
	
	Indeed, the automaton $A$ cannot come to the node in the state $q$
	twice in the accepting computation, 
	otherwise it will repeat a configuration and loop.
	By similar reasons the automaton cannot return to the initial node
	in the state $q_0$ in the accepting computation.
	
	Denote by $Q_{in}$ the set of all states in which the automaton $A$ visits the node,
	according to the information in the node label $(a,E)$.
	If $a \notin \Sigma_0$, then $Q_{in} = (\bigcup_{d \in D_a} Q_{in,d})$,
	if $a \in \Sigma_0$,
	then $Q_{in} = (\bigcup_{d \in D_a} Q_{in,d}) \cup \{q_0\}$.
\item
	For each state $q_1 \in Q_{in}$,
	either a transition $\delta(q_1,a)$
	or acceptance $(q_1,a) \in F$
	should be defined.
	If the transition $\delta(q_1,a) = (q_2,d)$ for some $q_2$ and $d$ is defined,
	then this transition should be encoded, that is $q_2 \in Q_{out,d}$ should hold.
	
	Indeed, if the automaton $A$ in the accepting computation
	visits some node in the state $q_1$, then it either accepts,
	or makes a transition, it cannot reject.
\item
	For each $d \in D_a$ and for each state $q_2 \in Q_{out,d}$,
	there must be a way to move from the current node in the state $q_2$ in the direction $d$.
	That is, there should exist a state $q_1 \in Q_{in}$,
	with $\delta(q_1,a) = (q_2,d)$.
\item
	For every two distinct states $p_1,q_1 \in Q_1$,
	with the transitions at the label $a$ defined,
	the transitions should be distinct:
	$\delta(p_1,a) \neq \delta(q_1,a)$.
	
	Indeed, the automaton in the accepting computation
	cannot come to the same configuration twice, otherwise it loops.
\end{enumerate}

The signature $S'$ has $k4^n$ directions.
There are at most $m4^{kn}$ node labels,
as in a label $(a,E)$ there are $m$ ways to choose an old label $a$,
and $4^n$ ways to choose sets $Q_{in,d}$ and $Q_{out,d}$ for each direction
$d \in D_a$.

All the directions with their opposite directions, and 
all the labels from $\widehat{\Sigma'}$ with their sets of directions
can be written down in time linear in their length,
that is, exponential in the length of the input.
Checking whether a label $(a,E) \in \widehat{\Sigma'}$
satisfies all conditions, can be done in linear time in the length of the label.

The signature $S'$ has been constructed, and it remains
to prove the correspondence between
the graphs over $S$ accepted by the automaton $A$,
and all the graphs over $S'$,
and to construct the functions $f$ and $g$ which define this correspondence.

Let the automaton $A$ accept some graph $G = (V,v_0,+,\lambda)$ over the signature $S$.
Then, the graph $f(G) = G' = (V',v_0',+,\lambda')$ over the signature $S'$ 
is constructed as follows.
\begin{itemize}
\item 
	The set of nodes and the initial node are the same: $V' = V$, $v_0' = v_0$.

\item
	The edges in $G'$ are the same as in $G$, but with additional information
	encoded in the directions.
	Let some edge $e$ with directions $(d,-d)$ connect the nodes $u$ and $v$ in the graph $G$,
	that is, $u+d = v$ in $G$. Let $Q_{in} \subseteq Q$ be a set of states
	in which the automaton $A$ in its computation comes to the node $u$ 
	from the node $v$ by the edge $e$, let $Q_{out} \subseteq Q$ be a set of states,
	in which the automaton $A$ arrives to the node $v$ from the node $u$ by the edge $e$.
	Then, the corresponding edge in $G'$ is defined by $u+(d,Q_{in},Q_{out}) = v$
	and $v+(-d,Q_{out},Q_{in}) = u$. These are all edges in $G'$.

\item
	The node labels in $G'$ are the node labels from $G$,
	but with added information on the new directions.
	Let a node $v$ in $G$ have label $a$,
	and accordingly edges in directions from $D_a$. 
	These directions in the graph $G'$ are augmented with the information about
	the automaton's moves, forming the set $E$ of new directions.
	Then the node label of the node $v$ in the graph $G'$ is $(a,E)$.
	The label $(a,E)$ is in $\Sigma'$,
	because it encodes the moves of the automaton
	in the accepting computation
	(and only labels encoding situations
	impossible in accepting computations were not included in $\Sigma'$).
	The node $v$ has edges in directions from $E = D'_{(a,E)}$.
	And only the initial node has an initial label, 
	because the new labels' being initial depends only on the component $a$ of $(a,E)$.
\end{itemize}

Now it remains to check, that each graph over $S'$ corresponds to some graph over $S$
that is accepted by $A$.

What is the general form of a graph $G'$ over $S'$?
In the first components 
of directions and node labels,
it encodes some graph $g(G') = G$ over $S$
(and this is a definition of $g$).
Then $g(f(G)) = G$ by definition.
The other components of directions and node labels 
encode some information about moves of the automaton.
It will be shown that all moves from the computation of the automaton $A$ on a graph $G$
must be encoded, and that looping or rejecting cannot be encoded.
Then, for each graph $G'$ over $S'$,
the corresponding graph $G=g(G')$ must be accepted by the automaton $A$.
Note that, besides all moves from the accepting computation, 
the graph $G'$ may additionally encode some cycles of transitions
that do not intersect with the accepting computation.
So an accepted graph $G$ over $S$ may have several pre-images $G'$,
such that $g(G') = G$.

It remains to prove that each graph $G'$ over the signature $S'$ 
must encode all moves the automaton $A$ makes in its computation
on the graph $G = g(G')$, and possibly some moves not in this computation, 
and that the computation of $A$ on $G$ must be accepting.

Fix a graph $G'$ over the signature $S'$, let $G=g(G')$,
and let $C = C_0, C_1, \ldots, C_N$ 
be the computation of the automaton $A$ on the graph $G$, 
where $C_N$ is the last configuration, or $N = \infty$ if the automaton loops.
It should be proved that $C$ is accepting and is encoded in $G'$.

This is proved by induction on $i$ 
that either the configuration $C_i$ is accepting,
or the next configuration $C_{i+1}$ exists and it is
different from all previous configurations, 
and the move from configuration $C_i$ to $C_{i+1}$ is encoded in $G'$.

Let $i \in \{0,1,2,\ldots, N\}$, and let the claim be proved for all $j < i$.

Denote the $i$-th configuration by $(q,v)$. Let $(a,E)$ be the label of the node $v$ in $G'$.
Then, one can define $Q_{in}$ for the label $(a,E)$ 
as in the conditions on $\Sigma'$.
If $i = 0$, then $a \in \Sigma_0$ and $q = q_0 \in Q_{in}$.
Otherwise, the move from $C_{i-1}$ to $C_i$ is encoded in $G'$,
and $q \in Q_{in}$ as well.
Then, by the second condition, as $q \in Q_{in}$, either $(q,a) \in F$, or $\delta(q,a) = (r,d)$,
for some $r \in Q$, $d \in D$,
and the transition is encoded as $r \in Q_{out,d}$.
In the latter case $r$ will be in $Q_{in,-d}$ for the node $v+d$.
It remains to check that $C_i$ is different from all previous configurations.
If $i = 0$, then this is true.
Now, let $(p,u)$ be the previous configuration, with $\delta(p,\lambda(u)) = (q,d)$
and with $u+d = v$. Then $q \in Q_{in,-d}$ for the label $(a,E)$ of the node $v$.
The first condition gives that the automaton could not have entered the node $v$ 
in the state $q$ from another direction earlier in the computation, 
and that $(q,v)$ cannot be the initial configuration.
And if the previous direction is the same, then the $4$-th condition prohibits
entering $(q,v)$ earlier from a previous state other than $p$.
Then, only $(p,u)$ can be the previous configuration
for $(q,v)$, and, by the induction hypothesis, $(p,u)$ is unique in $C_0, \ldots, C_{i-1}$.
Then, $(q,v)$ is unique in $C_0, \ldots, C_i$.

Thus, the computation of $A$ on $G$ is encoded in $G'$, this computation 
cannot loop, cannot reject, so it is accepting.
\end{proof}

Using Theorem~\ref{theorem_gwa_to_signature}
that reduces graph-walking automata
to signatures, one can solve the non-emptiness problem for graph-walking automata
in nondeterministic exponential time.

\begin{corollary}\label{theorem_gwa_non_emptiness_is_in_NEXP}
The problem of whether a given graph-walking automaton
accepts at least one graph is in NEXP.
\end{corollary}
\begin{proof}
First, the algorithm from Theorem~\ref{theorem_gwa_to_signature}
is applied to a given signature $S$ and to a given graph-walking automaton $A$ over this signature,
and it constructs a signature $S'$, such that there exist functions
$f \colon L(A) \to L(S')$ and $g \colon L(S') \to L(A)$.
Then, $L(A)$ is non-empty if and only if $L(S')$ is non-empty.
The size of the signature $S'$ is exponential in the size of $S$ and $A$,
and this signature is constructed in exponential time.
Checking whether $L(S')$ is non-empty can be done 
in nondeterministic polynomial time in the size of $S'$,
that is, in nondeterministic exponential time in the sum of sizes of $S$ and $A$.
\end{proof}

Actually, the non-emptiness problem for graph-walking automata
is NEXP-complete, that will be proved in Section~\ref{NP_completeness_for_signature_and_NEXP_for_gwa}.

An upper bound on the number of nodes in the minimal graph
accepted by a graph-walking automaton can be derived 
from the analogous bound for signatures.

\begin{corollary}\label{corollary_small_graphs_gwa}
Let $S = (D, -, \Sigma, \Sigma_0, (D_a)_{a \in \Sigma})$ be a signature
with $k \geqslant 2$ directions, with $m$ node labels,
and with $|D_a| \geqslant 1$ for each $a \in \Sigma$.
Let $A = (Q, q_0, F, \delta)$ be a graph-walking automaton over $S$ with $n$ states.
Then, if $A$ accepts at least one graph, then the number of nodes in
the smallest accepted graph is at most $m4^{n(k+1)}k^{k4^n-1}$. 
\end{corollary}
\begin{proof}
Let $A$ accept at least one graph. 
By Theorem~\ref{theorem_gwa_to_signature}, there is a signature $S'$,
and functions $f \colon L(A) \to L(S')$ 
and $g \colon L(S') \to L(A)$ that do not change the number of nodes in a graph. 
So the minimal number of nodes
for graphs over $S$ accepted by $A$ is equal to the minimal
number of nodes in graphs over $S'$.

The signature $S'$ has $k4^n$ directions, at most $m4^{kn}$ node labels,
and the maximum degree of a node at most $k$. The latter is because $f$ preserves
edge structure of graphs. Then, by Theorem~\ref{theorem_small_graphs_signature},
the minimal graph over the signature $S'$ has the number of nodes at most
\begin{equation*}
	m4^{nk}k4^n \min \Big\{\big(\frac{1}{2}k4^n\big)^{\frac{k4^n}{2}}, k^{k4^n-2}\Big\}
\leqslant m4^{nk}k4^n k^{k4^n-2} = m4^{n(k+1)}k^{k4^n-1}.
\end{equation*}
\end{proof}

\section{Computational complexity of emptiness problems} \label{NP_completeness_for_signature_and_NEXP_for_gwa}

It has been proved that the non-emptiness problems for signatures 
and for star automata are both in NP,
and that the non-emptiness problem for graph-walking automata is in NEXP.
In this section, all these problems are proved to be complete in their complexity classes.

NP-hardness of the non-emptiness problem for signatures
is proved by a reduction of graph 3-colourability to this problem.

\begin{theorem}\label{theorem_nonemptiness_signature_NP_complete}
The problem of whether there is at least one graph over a given signature
is NP-hard.
\end{theorem}
\begin{proof}

\begin{figure}[t]
	\centerline{\includegraphics[scale=0.9]{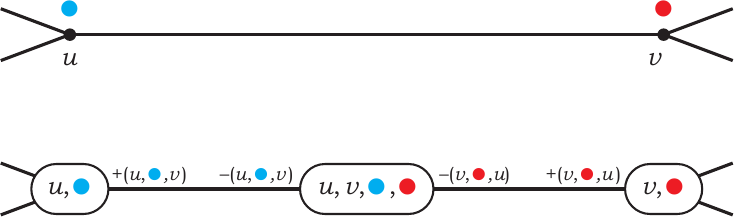}}
	\caption{From 3-colourability to signature non-emptiness:
		mapping a graph with a colouring
		to a graph over a signature.}
	\label{f:graph_signature_np_completeness}
\end{figure}

The 3-colourability problem for a connected graph $G = (V,E)$ is to check
whether its nodes can be coloured in $\{1, 2, 3\}$, so that every edge
connects differently coloured nodes.

For an input graph $G$, one should construct such a signature $S_G$ in polynomial time,
that there exists a graph over $S_G$ if and only if the graph $G$
can be coloured correctly.

The signature $S_G$ will be constructed so, that graphs over it
correspond to correct colourings of the graph $G=(V, E)$.

Nodes of $G$ can have any of the three colours,
and for each node and for each colour there is a corresponding node label.
Furthermore, for every edge with two distinct colours on its ends,
there is a separate node label representing this edge with these colours,
that is, an unordered pair of two coloured nodes.
\begin{equation*}
\Sigma = \setbig{(v,i)}{v \in V, \: i = 1,2,3} \cup
\setbig{\{(u, i), \, (v, j)\}}{(u,v) \in E, \: i,j \in \{1,2,3\}, i \neq j}.
\end{equation*}
The condition of the colouring to be correct
is checked by not having labels of the form $\{(u, i), \, (v, i)\}$,
representing edges with the same colour at both ends.

Fix any node $v_0 \in V$,
and let all labels $(v_0,i)$, with $i=1,2,3$, be initial.

The set of directions is organized so that
for every edge $(u, v)$ in the graph $G$,
node labels $(u,i)$ and $(v,j)$,
which correspond to the nodes $u$ and $v$ in the graph $G$,
would require a connection through an intermediate node
that corresponds to the edge $(u, v)$ in $G$,
and which gathers information on the colours of the nodes $u$ and $v$.
\begin{equation*}
D = \set{\pm(u,i,v)}{u,v \in V,\: (u,v) \in E, \: i = 1,2,3}.
\end{equation*}

The opposite direction to $+(u,i,v)$ is given by $-(u,i,v)$,
for all $u,v \in V$ with $(u,v) \in E$, and for all $i = 1,2,3$.

Each node of a graph over this signature which
represents one of the nodes of $G$ 
should be connected with the nodes representing all the edges coming out of this node.
\begin{equation*}
D_{(u,i)} = \set{+(u,i,v)}{v \in V, \: (u,v) \in E}, \text{ for all } u \in V,\: i = 1,2,3.
\end{equation*}
\begin{equation*}
D_{\{(u, i), \, (v, j)\}} = \{-(u,i,v), -(v,j,u)\}, \text{ for all } u,v \in V, \: (u,v) \in E, \: i,j \in \{1,2,3\}, \: i \neq j
\end{equation*}

It remains to prove that the signature $S_G$ is as desired,
that is, there is a graph over $S_G$ if and only if
there is a correct 3-colouring of $G$.

First of all, if a coloring $c \colon V \to \{1,2,3\}$ exists,
then a graph $G_c$ over $S_G$ representing this colouring
is constructed with the set of nodes $V \cup E$,
where each node $v \in V$ has label $(v, c(v))$,
each node $(u, v) \in E$ has label $\{(u, c(u)), \, (v, c(v))\}$.
For every edge $(u, v) \in E$ in the graph $G$,
the graph $G_c$ has edges from $u$ to $(u, v)$ and from $(u, v)$ to $v$,
with the appropriate directions,
as illustrated in Figure~\ref{f:graph_signature_np_completeness}.

Conversely, let $\widehat{G}$ be any graph over the signature $S_G$.
It is claimed that in this case there exists a correct 3-colouring of $G$,
and moreover, $\widehat{G}=G_c$ for some correct 3-colouring $c$ of $G$.

First, it is proved that for each node $v \in V$ of the graph $G$,
there is exactly one node in $\widehat{G}$ with a label of the form $(v,i)$, for some $i$.
Consider the shortest simple path from $v_0$ to $v$ in $G$
(it exists because $G$ is connected);
the proof is by induction on the length of this path.
The base case is a path of length 0:
here the node corresponding to $v_0$ exists because $\widehat{G}$ must have an initial node,
and it is unique because the initial node is unique.
For the induction step, let $u$ be the next to the last node on the path, with $(u, v) \in E$.
By the induction hypothesis, in $\widehat{G}$,
there is a unique node of the form $(u, i)$, for some $i$.
This node emits a unique edge in the direction $+(u, i, v)$,
which must lead to a node labelled with $\{(u, i), \, (v, j)\}$, for some $j$,
which in turn emits a unique edge in the direction $-(v, j, u)$
that ends in a node labelled with $(v, j)$---so this node exists.
If there were another node in $\widehat{G}$ labelled with $(v, k)$, for any $k$,
then, by the same reasoning, it would be connected to some node labelled with $(u, \ell)$
through some intermediate node;
this node must be the same as the above node labelled with $(u, i)$,
because such a node is unique.
However, there is a unique path simulating the edge $(u, v)$,
hence this node labelled with $(v, k)$ must coincide with the above node labelled with $(v, j)$.

Therefore, $\widehat{G}$ has the set of nodes $V \cup E$,
which replicates the structure of $G$,
with every edge split by an intermediate node.
Then, it must be $G_c$ for some colouring $c$.
This colouring is correct, because each intermediate node
checks that the colours at both ends of the corresponding edge are distinct.
Then, correct colourings of the graph $G$ correspond to graphs over $S_G$.

Note that the intermediate nodes that split the edges of $G$ are necessary,
because node labels cannot accumulate information on the colours of all the neighbours of a node,
as this would require an exponential number of node labels.
\end{proof}

The non-emptiness problem for star automata is NP-complete as well.
Its membership in NP was established above,
and its NP-hardness follows from the NP-hardness of non-emptiness of signatures.

\begin{theorem}\label{theorem_nonemptiness_star_NP_complete}
The problem of checking whether a given star automaton accepts at least one graph
is NP-hard.
\end{theorem}
\begin{proof}
Non-emptiness for signatures was proved in Theorem~\ref{theorem_nonemptiness_signature_NP_complete} to be NP-hard.
Now the NP-hardness of the non-emptiness problem for star automata
is proved by reducing the non-emptiness problem for signatures to it, as follows.

Let $S$ be a given signature. Consider the automaton $A_*$ over it,
that has one state, and,
for each node label, has a star with
this state at the centre and with this state at all rays.
This star automaton accepts all graphs, so its non-emptiness is equivalent to
non-emptiness of the signature $S$. And this automaton $A_*$ has size polynomial
in the size of $S$.
\end{proof}

Now it is time to prove the NEXP-completeness of the non-emptiness problem for
graph-walking automata.
It was proved in Corollary~\ref{theorem_gwa_non_emptiness_is_in_NEXP}, 
that this problem is in NEXP.
For NEXP-hardness it will be proved that a signature and a graph-walking automaton
can define a set of graphs containing a square grid of size exponential 
in the number of states of the automaton and in the size of the signature.
And then a nondeterministic Turing machine working in exponential time
will be simulated on such grids.

\begin{theorem}\label{theorem_nonemptiness_gwa_NEXP_complete}
The problem of whether there is at least one graph accepted 
by a given graph-walking automaton is NEXP-hard.
\end{theorem}
\begin{proof}
Fix some NEXP-complete problem and some nondeterministic Turing machine $M$
that solves this problem in exponential time. It can be assumed that
the Turing machine is one-tape with the tape infinite to the right,
and that the machine never moves to the left from the first position of the tape,
in which an input string begins.
The number of states, the number of transitions in the transition function,
the sizes of the input aphabet and of the work alphabet are constant, 
as the Turing machine $M$ is fixed.

The problem whether a given string $w$ over the input alphabet is accepted 
by the Turing machine $M$ is NEXP-complete.
So to prove the theorem it is enough to reduce in polynomial time 
this problem about $M$
to the non-emptiness problem for a graph-walking automaton.
That is, such a deterministic polynomial-time algorithm is needed, 
that for a given string $w$ it constructs 
a signature $S_w$ and a graph-walking automaton $A_w$
so that a graph accepted by $A_w$ will exist if and only if
there exists an accepting computation of the machine $M$ on the string $w$.

Let $f \colon \mathbb{N} \to \mathbb{N}$ be a polynomial-time computable function
that, for each length $\ell$ of an input string, gives a number $f(\ell)$, 
bounded by a polynomial in $\ell$, 
such that $f(\ell) \geqslant \max\{\ell,2\}$, and that the Turing machine $M$ halts on
every string of length at most $\ell$ in not more than $2^{f(\ell)}-1$ steps. 
Then, each computation of $M$ on each string of length at most $\ell$
can be written on a grid of length $2^{f(\ell)}\times 2^{f(\ell)}$.

The signature $S_w = (D, -, \Sigma, \Sigma_0, (D_a)_{a \in \Sigma})$
depends only on the length of $w$ and is constructed as follows.

Let $n = f(|w|)$, so that each computation of $M$ on a string $w$ 
can be written on a grid of size $2^n \times 2^n$; the number $n = f(|w|)$ 
can be computed in polynomial time and is polynomial in the length of $w$.

The signature $S_w$ is composed of three parts: $D = D_1 \cup D_2 \cup D_3$, 
$\Sigma = \Sigma_1 \cup \Sigma_2 \cup \Sigma_3$, all sets here are disjoint.
And for each node label $a \in \Sigma_i$, it should hold that $D_a \subseteq D_i$,
for $i \in \{1,2,3\}$.
In every graph over $S_w$ all nodes are divided into three sets: $V = V_1 \cup V_2 \cup V_3$,
where $V_i$ consists of the nodes with labels in $\Sigma_i$, for $i = 1,2,3$.
There are two special pairs of opposite directions: $+d \in D_1$ and $-d \in D_2$,
and $+d' \in D_2$ and $-d' \in D_3$.
For every other direction, the opposite direction lies in the same set.
Thus, nodes in $V_1$ and nodes in $V_2$ can be connected only by $(+d,-d)$-edges;
similarly, nodes in $V_2$ and nodes in $V_3$
can be connected only by $(+d',-d')$-edges. A node from $V_1$
and a node from $V_3$ cannot be connected with an edge.

The idea is that nodes with labels in $\Sigma_2$ form a grid 
on which the Turing machine working on $w$ will be simulated. 
Each node label from $\Sigma_2$ will have both directions $-d$ and $+d'$.
Labels from $\Sigma_1$ will allow the nodes in $V_1$ to form only 
a full binary tree of height $2n$ that
emits exactly $2^{2n}$ edges in the direction $+d$ from its leaves, 
thus ensuring that in every graph the number of nodes in $V_2$ is exactly $2^{2n}$.
Labels from $\Sigma_3$ will be used to attach a chain of length $2n$ 
to every node with label in $\Sigma_2$, with the chain consisting 
of zeros and ones. 
The automaton $A_w$ will check that nodes in $V_2$ form a $2^n \times 2^n$ grid,
and that chains attached to these nodes 
correctly encode the row number and the column number in the grid for each node.
Next, the automaton $A_w$ will check that
some accepting computation of the Turing machine $M$
on the string $w$ is encoded on the grid.
Figure~\ref{f:gwa_emptiness_nexp} shows a graph over some signature $S_w$ with $n = 2$,
that defines a correct grid on nodes with labels in $\Sigma_2$.

The only initial node label in the signature $S_w$ is $a_0 \in \Sigma_1$.
The first part $\Sigma_1$ and $D_1$ should be defined 
so that the nodes with labels in $\Sigma_1$ 
can form only one graph: a full binary tree of height $2n$ with $2^{2n}$ leaves.
The set of node labels is 
$\Sigma_1 = \{a_0, a_1, b_1, a_2, b_2, \ldots, a_{2n}, b_{2n}\}$,
and the set of directions is 
$D_1 = \{\pm \ell_1, \pm r_1, \pm \ell_2, \pm r_2, \ldots, \pm \ell_{2n}, \pm r_{2n}\} \cup \{+d\}$.
Here the label $a_0$ is initial, it is used for the root of a tree (level $0$),
the labels $a_i$ and $b_i$ are used for left and right children of
the $i$-th level.
The node label $a_0$ has the set of directions $D_{a_0} = \{+\ell_1, +r_1\}$,
that is, the root has two edges to the two nodes of level $1$.
Labels $a_i$ and $b_i$, for $i \in \{1, \ldots, 2n-1\}$,
have the sets of directions $D_{a_i} = \{-\ell_i,+\ell_{i+1},+r_{i+1}\}$ and 
$D_{b_i} = \{-r_i,+\ell_{i+1},+r_{i+1}\}$. So the $i$-th level
generates twice as many nodes on level $i+1$.
The node labels of the last level $2n$ (for the leaves of the tree) have sets of directions 
$D_{a_{2n}} = \{-\ell_{2n}, +d\}$ and
$D_{b_{2n}} = \{-r_{2n},+d\}$, that is, each leaf emits one edge
in the direction $+d$, which is used for connection with nodes in $V_2$.

Thus, in every graph over the signature $S$ the initial node is labelled with $a_0 \in \Sigma_1$
and all nodes in $V_1$ form a full binary tree with $2^{2n}$ leaves and each leaf emits
an edge in the direction $+d$.

The part $\Sigma_3$, $D_3$ is constructed to allow only chains of nodes of length $2n$
with one direction $-d'$ in each chain, with zeros and ones in nodes.
This part of the signature is defined by
$\Sigma_3 = \{0_1,\dots, 0_{2n}\} \cup \{1_1,\dots, 1_{2n}\}$, 
$D_3 = \{\pm d_1, \ldots, \pm d_{2n-1}\} \cup \{-d'\}$.
And $D_{0_1} = D_{1_1} = \{-d', +d_1\}$; 
$D_{0_i} = D_{1_i} = \{-d_{i-1},+d_i\}$, for $i \in \{1, \ldots, 2n-1\}$; and
$D_{0_{2n}} = D_{1_{2n}} = \{-d_{2n-1}\}$.

Then, each node in $V_2$ has a chain attached to it in the direction $+d'$.
Every such chain consists of nodes with labels in $\Sigma_3$, has length $2n$
and encodes a number from $0$ to $2^{2n}-1$ in a sequence of zeros and ones
in nodes.
Let some node $v$ in a graph have a label in $\Sigma_2$.
Then, the \emph{coordinates} of $v$ are the pair of numbers $(i_v,j_v)$, 
for $i_v, j_v \in \{0, \ldots, 2^n-1\}$, where the number $i_v$
is defined by the first $n$ bits in the chain of nodes in $V_3$ attached to $v$,
and the number $j_v$ is defined by the second $n$ bits.
The number $i_v$ is meant to be the number of the row in the grid where $v$ is located, 
and $j_v$ is meant to be the number of the column.
Note that the coordinates of the node $v$ are by definition just a pair of numbers, 
encoded in a chain,
even if these numbers do not correspond to the actual position of the node $v$ in a grid.

\begin{figure}[p]
	\centerline{\includegraphics[scale=0.9]{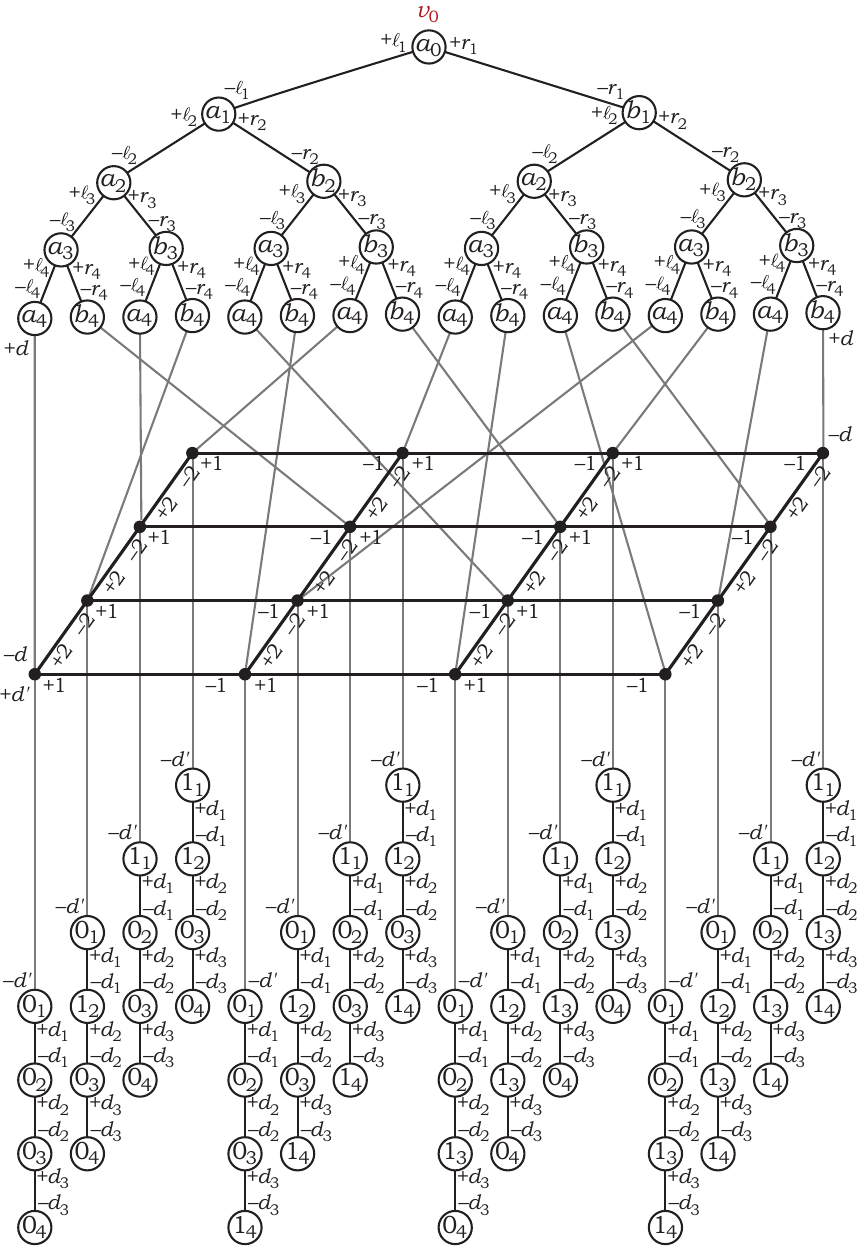}}
	\caption{A graph that defines a correct $2^n \times 2^n$ grid, for $n = 2$.
	The graph has three levels: the tree on the nodes in $V_1$ at the top,
	the grid on the nodes in $V_2$ in the middle,
	and chains on nodes in $V_3$ at the bottom, which encode the coordinates of nodes in the grid
	(the upper two bits encode the row number, and the lower two bits
	encode the column number).}
	\label{f:gwa_emptiness_nexp}
\end{figure}

Now to the main part of the signature: $\Sigma_2$ and $D_2$. There are $6$
directions in $D_2$: two of them, $-d$ and $+d'$, are used for connection with $V_1$ and $V_3$,
and $4$ directions are used for a grid: $\pm 1$ are horizontal 
($+1$ is right, $-1$ is left), and $\pm 2$ are vertical
($+2$ is up, $-2$ is down), so $D_2 = \{\pm 1, \pm 2\} \cup \{-d,+d'\}$.
The set of node labels is $\Sigma_2 = \text{Pos} \times \text{Alph} \times \text{Head}$,
that is, each node label in $\Sigma_2$ is of the form 
$(\text{\emph{pos}}, \text{\emph{alph}}, \text{\emph{head}})$.
The first component \emph{pos} gives the type of a position of a node in a grid:
in one of $4$ corners, on the side or in the centre.
So there are $9$ variants of the first component of a node label: 
\begin{align*}
\text{Pos} = 	\{ &\text{LU}, \text{CU}, \text{RU}, \\
			    &\text{LC}, \text{CC}, \text{RC}, \\
			    &\text{LD}, \text{CD}, \text{RD}\},
\end{align*}
where the first letter of \emph{pos} gives the type of horizontal position ($L$, $C$ or $R$), 
and the second letter gives the type of vertical position ($D$, $C$ or $U$).
The set of directions $D_a$ for each node label $a \in \Sigma_2$ 
depends only on the component \emph{pos} of the label $a$: the directions $-d$ and $+d'$
are always in $D_a$; the direction $+1$ is in $D_a$ if the node is not at the right border of a grid,
that is, if $\text{\emph{pos}} \notin \{RU, RC, RD\}$; the direction $-1$ is in $D_a$ if and only if
$\text{\emph{pos}} \notin \{LU, LC, LD\}$; similarly $+2 \in D_a$ if and only if 
$\text{\emph{pos}} \notin \{LU, CU, RU\}$; and $-2 \in D_a$ if and only if 
$\text{\emph{pos}} \notin \{LD, CD, RD\}$.

The components \emph{alph} and \emph{head} of node labels in $\Sigma_2$ will be used
for simulating configurations of the Turing machine $M$ on rows of a grid.
Let $\Gamma$ be the work alphabet of $M$,
it contains the input alphabet, the new blank symbol and maybe some other symbols; 
let $Q$ be a finite set of states of the Turing machine.
Then, $\text{Alph} = \Gamma$, that is, the component \emph{alph} gives one
of the symbols in the work alphabet of $M$, and 
$\text{Head} = Q \cup \{0\}$, where $0 \notin Q$,
that is, the component \emph{head} gives either a state of the Turing machine $M$
if the head is simulated at the current position, or $\text{\emph{head}} = 0$ 
if there is no head in this position.

This signature $S_w$ is constructed in time linear in $n$.

Now a graph-walking automaton $A_w$ over the signature $S_w$ should be constructed,
so that it accepts only graphs, in which nodes in $V_2$ form a correct grid, and
the components \emph{alph} and \emph{head} of the labels in these nodes encode
a correct accepting computation of the Turing machine $M$ on the string $w$.
The work of the automaton $A_w$ on a graph is divided into two phases:
checking the grid and checking the encoding of the Turing machine's computation on that grid.

In the first phase the automaton does not distinguish the components \emph{alph} and \emph{head} 
in labels in $\Sigma_2$,
its actions on a node labelled with 
$(\text{\emph{pos}},\text{\emph{alph}},\text{\emph{head}}) \in \Sigma_2$
depend only on the component \emph{pos}.

The goal of the first phase is to check that nodes with labels in $\Sigma_2$
form a $2^n \times 2^n$ grid on directions $\pm 1$ and $\pm 2$,
and that the coordinates $(i_v,j_v)$ of each node $v$ in $V_2$ 
are numbers of its row and its column in a grid.
For convenience, the automaton also checks
that the leftmost path in the tree on nodes in $V_1$ leads to a node in $V_2$
with coordinates $(0,0)$.
If all these conditions hold for a graph, then this graph is said to define a correct grid.

The automaton checks whether a graph defines a correct grid as follows.
\begin{enumerate}
\item
	At the beginning, the automaton checks that the leftmost path in the tree on nodes with
	labels in $\Sigma_1$ leads to a node with a label in $\Sigma_2$
	that has coordinates $(0,0)$. The automaton starts at the initial node at the root of the tree,
	then it moves to the left child until it comes to a node with label in $\Sigma_2$.
	Then it checks that all nodes in the attached chain contain zeros.
	This can be done with a constant number of states.
\item
	The automaton checks, for each node $v$ with label in $\Sigma_2$, that
	the component \emph{pos} of the label agrees with coordinates $(i_v,j_v)$
	given in the chain of nodes from $V_3$ attached to the node $v$.
	For that, it should be checked that $\text{\emph{pos}} = XY$, 
	where
	\begin{equation*}
		X = 
	 	\begin{cases}
	   		L & \text{if } j_v = 0\\
	   		C & \text{if } 0 < j_v < 2^n-1\\
	   		R & \text{if } j_v = 2^n-1
	 	\end{cases}
	 	\hspace{2cm}
		Y = 
	 	\begin{cases}
	   		D & \text{if } i_v = 0\\
	   		C & \text{if } 0 < i_v < 2^n-1\\
	   		U & \text{if } i_v = 2^n-1
	 	\end{cases}
	\end{equation*}.
	
	When the automaton visits some node $v \in V_2$, it can check this condition
	for the node $v$ using a constant number of states and return to the node.
	Indeed, it needs just to check several conditions of the form 
	that all bits of the first or the second $n$ bits of a chain are all zeros
	or are all ones.
	
	To do such a check for each node in $V_2$, the automaton needs to visit 
	somehow all nodes in $V_2$.
	This can be done by traversing the tree on nodes in $V_1$. 
	This tree can be traversed with a constant number of states. The leaves in this tree
	correspond to nodes in $V_2$, 
	each leaf is connected by a $(+d,-d)$ edge to some node in $V_2$, and
	each node in $V_2$ is connected to some leaf.
	Thus, the automaton checks for each leaf in a tree
	that its neighbour in $V_2$ has the component \emph{pos} agree with the coordinates. 
	This can be done using a constant number of states.
\item
	Then the automaton checks that directions $\pm1, \pm 2$ in the grid 
	lead to correct nodes. That is, for each node $v$ in $V_2$ 
	with coordinates $(i_v,j_v)$, the following conditions must hold.
	If an edge in the direction $-1$ exists ($j_v > 0$),
	then it should lead to a node with coordinates $(i_v,j_v-1)$.
	If an edge by $+1$ exists ($j_v < 2^n-1$), then
	it should lead to a node with coordinates $(i_v,j_v+1)$.
	Similarly, the direction $+2$ must increase the coordinate $i_v$,
	and the direction $-2$ must decrease it.
	
	When the automaton visits some node $v$ in $V_2$, it can check these conditions
	using $O(n)$ states and return to the node $v$. Indeed, to check
	the equality of two vectors of length $n$ contaning $0$s and $1$s,
	the automaton can compare them bit by bit
	remembering only the position of the current bit in a vector and the value of this bit.
	To check that the number encoded in the first vector is greater by $1$ than
	the number encoded in the second vector,
	the automaton can check that the binary representations of the vectors
	are of the form $x10^i$ and $x01^i$, with $x \in \{0,1\}^*$ and $i \geqslant 0$,
	and this can be checked bit by bit.
	
	To make these checks for all nodes in $V_2$ the automaton traverses the tree
	on the nodes in $V_1$ as at the previous step.
\item
	If the automaton did not reject at the previous steps, 
	then it returns to the node with coordinates $(0,0)$ and starts the second phase. 
\end{enumerate}

If the automaton rejects at the first phase, then the graph does not define a correct grid.
It is claimed that the checks the automaton makes are sufficient, that is,
that if the automaton starts the second phase, then the graph defines a correct grid.
Let the automaton start the second phase on some graph $G$.

First, it is shown that all nodes with labels in $\Sigma_2$ have distinct coordinates
and that every pair of coordinates $(i,j)$, for $i,j \in \{0,\ldots,2^n-1\}$, occurs somewhere.

The node with coordinates $(0,0)$ exists because such a node is on the leftmost path.
For each node with some coordinates
$(i,j)$, the automaton has checked that its neighbours in directions $\pm 1, \pm 2$
exist and have coordinates $(i,j+1)$, $(i,j-1)$, $(i+1, j)$, $(i-1,j)$,
as long as these coordinates are between $0$ and $2^n-1$. 
Then, for all $i,j \in \{0,\ldots,2^n-1\}$, there is a node in $V_2$ with
coordinates $(i,j)$.
As the tree on nodes in $V_1$ is defined uniquely, $|V_2| = 2^{2n}$ in every graph.
So the node with each pair of coordinates is unique.

Note that the automaton has no way to distinguish a node from its copy locally,
so it is important that counting arguments give uniqueness
to each pair of coordinates.

Then, as a node with each pair of coordinates exists and is unique, and 
coordinates increase or decrease along the directions in the grid,
the graph defines a correct grid.

The states and transitions used by the automaton in the first phase
can be written down in time quadratic in $n$, as both the number of states
and the number of node labels in the signature are linear in $n$.

In the second phase, the automaton checks that some accepting computation
of the Turing machine $M$ on the string $w$ is encoded in the grid. 

The automaton should check that the initial row encodes
the initial configuration of the Turing machine $M$ on the string $w$,
that the next row encodes one of possible next configurations, and so on,
up to an accepting configuration. Rows after the accepting configuration
are allowed to contain anything.

How are configurations encoded in rows?
The Turing machine $M$ works in exponential time, and the number $n$ was chosen so
that every computation on $w$ contains at most $2^n-1$ steps,
and that $|w| \leqslant n$. Thus, the head of the Turing machine never visits
positions beyond $2^n-1$ on the tape,
and during the computation the symbols at these positions
are blank symbols. So the tape contents in a configuration can be thought of as a string
of length $2^n$. This string is encoded in the nodes of a row in the
components \emph{alph} of node labels, one symbol of the string per node.
The position of the head is encoded by having the component \emph{head} 
non-zero only in one node;
in this node, the component \emph{head} encodes a state of the Turing machine. 

The automaton works in the second phase as follows.

\begin{enumerate}
\item 
	The automaton $A_w$ starts the second phase on a graph 
	at the node with coordinates $(0,0)$,
	and the graph is known to define a correct grid of size $2^n \times 2^n$.

\item
	First, the automaton checks the encoding of the initial configuration.
	It goes through the first $|w|$ nodes in the first row 
	remembering in a state the number of moves $j$ it made,
	and for each node it checks that the component \emph{alph} of the node label 
	is the $j$-th symbol of $w$.
	Then it continues moving to the right using one state for that, 
	and checking that the components \emph{alph} in all other nodes in the first row
	contain blank symbols. 
	While moving from $(0,0)$ to $(0, 2^n-1)$ the automaton additionally checks
	that in the node $(0,0)$ the component \emph{head} contains 
	one of the initial states of the Turing machine,
	and that in all other nodes of the first row the component \emph{head} of the label is $0$.
\item
	For each row $i \in \{0, \ldots, 2^n-1\}$, 
	starting from the row $i = 0$, the automaton makes the following two actions.
	
	First, the automaton checks whether the current configuration is accepting.
	It finds the node in which the head is encoded,
	and if $(\text{\emph{head}}, \text{\emph{alph}})$ is an accepting pair of $M$,
	then the automaton immediately accepts.
	
	If the configuration encoded in the $i$-th row is not accepting, then the automaton
	checks that the next row encodes one of the possible next configurations.
	This check can be done using a constant number of states as follows.
	In the neighbourhood of the head in the $i$-th row,
	the automaton checks that a transition is correctly made;
	elsewhere, the automaton checks that the tape symbols are unchanged,
	and no extra heads appear.
	Once the check is complete, the automaton moves to the next row.
\end{enumerate}

Working as described above, the automaton accepts a graph in the second phase
if and only if one of the accepting computations of $M$ on $w$ is encoded on the grid,
and otherwise it rejects.
The automaton $A_w$ can be constructed in time polynomial in $n$,
and the NEXP-complete problem of whether the Turing machine $M$ accepts
a given string $w$ or not is reduced to the problem of
whether the graph-walking automaton $A_w$ over $S_w$ accepts at least one graph.
Thus, non-emptiness for graph-walking automata is NEXP-hard.
\end{proof}

\section{Conclusion}

In this paper it has been shown that the emptiness problems
for signatures, for star automata and for graph-walking automata are decidable.
And the computational complexity classes for these problems were determined:
the non-emptiness problems for signatures and for star automata are NP-complete,
whereas non-emptiness for graph-walking automata is NEXP-complete.
Table~\ref{t:emptiness_complexity} compares these new results about 
automata on graphs 
with the previous results for similar automata on strings and on trees.

Note that the reduction of graph-walking automata to signatures
works even in the case of nondeterministic graph-walking automata.
In this case, the conditions on incoming and outgoing states encoded in a label
should be replaced with
the conditions that the incoming states are all different,
and that there is a bijection
between the incoming and the outgoing states,
with a transition possible for each pair.
So the non-emptiness for nondeterministic graph-walking automata is NEXP-complete as well.

\begin{table}[h]
\centerline{\begin{tabular}{|c|c|c|c|}
\hline
	& strings
		& trees
			& graphs
				\\
\hline
walking
	& (2DFA)
		& (DTWA)
			& (DGWA)
				\\
	& PSPACE-complete {\small \cite{{Kozen}}}
		& EXP-complete~\cite{Bojanczyk}
		 	& NEXP-complete {\small (Cor~\ref{theorem_gwa_non_emptiness_is_in_NEXP}, Thm~\ref{theorem_nonemptiness_gwa_NEXP_complete})}
				\\
\hline
tilings by
	& (NFA)
		& (tree automata)
			& (star automata)
				\\
edges/stars
	& NL-complete {\small \cite{{Jones}}}
		& P-complete {\small \cite{Veanes}}
			& NP-complete {\small (Cor~\ref{theorem_nonemptiness_star_is_in_NP}, Thm~\ref{theorem_nonemptiness_star_NP_complete})}
				\\
\hline
\end{tabular}}
\caption{Complexity of the non-emptiness problem for different families of automata.}
\label{t:emptiness_complexity}
\end{table}

In this paper, several upper bounds on the number of nodes in minimal accepted graphs
have been obtained. Bounds have been proved 
for graph-walking automata (Corollary~\ref{corollary_small_graphs_gwa}), 
for star automata (Corollary~\ref{corollary_small_graphs_star_automata}),
and simply for graphs over a signature (Theorem~\ref{theorem_small_graphs_signature}).
It will be good to prove some lower bounds,
and maybe to improve the upper bounds given in this paper.

Star automata in this paper are a special case 
of elementary acceptors of Thomas~\cite{Thomas_tilings}, 
they are elementary acceptors without conditions on the number of occurrences of each star. 
Is the emptiness problem for elementary acceptors of Thomas
also decidable? This remains an open question.

\section*{Acknowledgements}

I am grateful to Alexander Okhotin for his advices on the presentation
and for helping to translate the paper to English.

I wish to thank Anton Martynov for suggesting to use Hadamard's bound on the determinants.


\begin{thebibliography}{99}

\bibitem{Bojanczyk} M. Boja\'nczyk,
	\href{https://doi.org/10.1007/978-3-540-88282-4_1}
	{``Tree-walking automata''},
	LATA 2008,
	LNCS 5196, 1--2.
	Extended version available at \url{https://www.mimuw.edu.pl/~bojan/upload/conflataBojanczyk08.pdf}.
	
\bibitem{BojanczykColcombet_det} M. Boja\'nczyk, T. Colcombet,
	\href{http://dx.doi.org/10.1016/j.tcs.2005.10.031}
	{``Tree-walking automata cannot be determinized''},
	\emph{Theoretical Computer Science},
	350:2--3 (2006), 164--173.
	
\bibitem{Budach} L. Budach,
	\href{http://dx.doi.org/10.1002/mana.19780860120}
	{``Automata and labyrinths''},
	\emph{Mathematische Nachrichten},
	86:1 (1978), 195--282.

\bibitem{FraigniaudIlcinkasPeerPelcPeleg} P. Fraigniaud, D. Ilcinkas, G. Peer, A. Pelc, D. Peleg,
	\href{http://dx.doi.org/10.1016/j.tcs.2005.07.014}
	{``Graph exploration by a finite automaton''},
	\emph{Theoretical Computer Science},
	345:2--3 (2005), 331--344.	
	
\bibitem{Hadamard} J. Hadamard,
	{``R\'esolution d'une question relative aux d\'eterminants''},
	\emph{Bulletin des Sciences Math\'ematiques},
	17 (1893), 240--246.

\bibitem{Jones} N. D. Jones,
	\href{http://dx.doi.org/10.1016/S0022-0000(75)80050-X}
	{``Space bounded reducibility among combinatorial problems''},
	\emph{Journal of Computer and System Sciences},
	11:1 (1975), 68--85.

\bibitem{Kozen} D. Kozen,
	\href{http://dx.doi.org/10.1109/SFCS.1977.16}
	{``Lower bounds for natural proof systems''},
	\emph{FOCS 1977}, 254--266.

\bibitem{KuncOkhotin_reversible} M. Kunc, A. Okhotin,
	\href{https://doi.org/10.1016/j.ic.2020.104631}
	{``Reversibility of computations in graph-walking automata''},
	\emph{Information and Computation},
	275 (2020), article 104631.
	
\bibitem{MartynovaOkhotin_lb} O. Martynova, A. Okhotin,
	\href{https://doi.org/10.4230/LIPIcs.STACS.2021.52}
	{``Lower bounds for graph-walking automata''},
	\emph{38th Annual Symposium on Theoretical Aspects of Computer Science}
	(STACS 2021, Saarbr\"ucken, Germany, 16--19 March 2021),
	LIPIcs 187, 52:1--52:13.

\bibitem{MartynovaOkhotin_gwa_boolean} O. Martynova, A. Okhotin,
	\href{https://doi.org/10.1007/978-3-030-93489-7_11}
	{``State complexity of union and intersection on graph-walking automata''},
	\emph{Descriptional Complexity of Formal Systems 2021},
	LNCS 13037, 125--136.

\bibitem{Rollik} H. A. Rollik,
	\href{https://doi.org/10.1007/BF00288647}
	{``Automaten in planaren Graphen''},
	\emph{Acta Informatica},
	13:3 (1980), 287--298.
	
\bibitem{Thomas_tilings} W. Thomas,
	\href{http://dx.doi.org/10.1007/3-540-54233-7_154}
	{``On logics, tilings, and automata''},
	\emph{Automata, Languages and Programming}
	(ICALP 1991, Madrid, Spain, 8--12 July 1991),
	LNCS 510, 441--454.

\bibitem{Veanes} M. Veanes,
	{``On computational complexity of basic decision problems of finite tree automata''},
	Technical Report 133, Uppsala University, Computing Science Department,
	1997.

\end{thebibliography}
\end{document}